\documentclass[12pt, a4paper]{article}

\usepackage[total={39pc, 55.5pc}, top=30mm, centering]{geometry}

\usepackage{setspace}
\usepackage{amsmath, amsthm, amssymb, amsfonts, mathrsfs, amsbsy, bm}

\usepackage{graphicx}
\usepackage{xcolor} 

\usepackage{algorithm}
\usepackage{algpseudocode}

\usepackage[nosort]{cite}

\usepackage[hidelinks]{hyperref}

\newcommand{\bx}{\mathbf{x}}
\newcommand{\by}{\mathbf{y}}

\newcommand{\bbC}{\mathbb{C}}
\newcommand{\bbR}{\mathbb{R}}

\newcommand{\calF}{\mathcal{F}}
\newcommand{\calJ}{\mathcal{J}}
\newcommand{\calK}{\mathcal{K}}

\newcommand{\calO}{\mathcal{O}}

\newcommand{\calR}{\mathcal{R}}

\newcommand{\ReS}{\mathrm{Re}\,S}
\newcommand{\ImS}{\mathrm{Im}\,S}

\newcommand{\vev}[1]{\langle #1 \rangle}

\newcommand{\tr}{\mathrm{tr}\,}
\newcommand{\re}{\mathrm{Re}\,}

\newcommand{\with}{{~~\mbox{with}~~}}

\makeatletter
\@addtoreset{equation}{section}

\def\@seccntformat#1{\csname the#1\endcsname.~~}
\makeatother

\begin{document}

\begin{titlepage} 

\renewcommand{\thefootnote}{\fnsymbol{footnote}}
\begin{flushright}
  KUNS-3067
\end{flushright}
\vspace*{1.0cm}

\begin{center}
{\Large \bf
Enhancing the ergodicity of Worldvolume HMC via embedding generalized thimble HMC
}
\vspace{1.0cm}

\centerline{
  {Masafumi Fukuma${}^1$}%
  \footnote{
    E-mail address: fukuma@gauge.scphys.kyoto-u.ac.jp
  }
  and 
  {Yusuke Namekawa${}^2$}%
  \footnote{
    E-mail address: namekawa@fukuyama-u.ac.jp
  }
}

\vskip 0.2cm
${}^1${\it Department of Physics, Kyoto University,
  Kyoto 606-8502, Japan}
\vskip 0.0cm
${}^2${\it Department of Computer Science, Fukuyama University,
  Hiroshima 729-0292, Japan}
\vskip 0.5cm

\end{center}

\begin{abstract}

The Worldvolume Hybrid Monte Carlo (WV-HMC) method [arXiv:2012.08468] is an efficient and versatile algorithm that mitigates the sign problem while resolving the ergodicity issues inherent in Lefschetz-thimble approaches. We focus on cases where the maximum flow time can be kept small, such as when applying WV-HMC to the doped Hubbard model utilizing a redundant, nonphysical parameter. An optimal choice of this parameter significantly reduces the sign problem on the original integration surface. This allows for small flow times, thereby enabling the simulation of larger system sizes at a modest computational cost. However, when the worldvolume reduces to a thin layer, phase-space exploration becomes inefficient, and ergodicity problems may reemerge. To address this limitation in WV-HMC, we propose embedding generalized thimble HMC (GT-HMC) into the WV-HMC framework. GT-HMC performs updates on a single deformed surface at a fixed flow time. Despite its inherent ergodicity issues at the zeros of the Boltzmann weight, GT-HMC efficiently explores the allowed region and typically permits larger molecular dynamics step sizes than WV-HMC. Consequently, it is highly effective in regions where ergodicity issues are less severe. We prove that GT-HMC can be consistently embedded within WV-HMC and confirm that the standalone and combined algorithms agree within statistical errors for the two-dimensional doped Hubbard model on an $8 \times 8$ lattice. This combined algorithm enables simulations on larger spacetime lattices. We demonstrate the feasibility of this approach by extrapolating the number and energy densities to the zero Trotter step limit at fixed temperature $T/t = 1/6.4\simeq 0.156$ and repulsive interaction $U/t = 8.0$. Even with modest sample sizes, we achieve controlled statistical errors across the entire range of the chemical potential.

\end{abstract}
\end{titlepage}

\pagestyle{empty}
\pagestyle{plain}

\tableofcontents
\setcounter{footnote}{0}

\section{Introduction}
\label{sec:intro}

The numerical sign problem is a major obstacle 
to first-principles computations 
of a wide range of physically important systems, 
including quantum chromodynamics (QCD) at finite density, 
strongly correlated electron systems, 
and real-time dynamics of quantum many-body systems. 
This problem arises when the action becomes complex, 
causing the Boltzmann weight to lose its positive-definite nature 
and become oscillatory. 
In systems with a large number of degrees of freedom, 
these oscillations lead to severe cancellations, 
making the numerical evaluation of observables exponentially difficult. 

Among recent attempts to find versatile solutions, 
the Lefschetz thimble method has attracted considerable attention 
\cite{Witten:2010cx,Cristoforetti:2012su,Cristoforetti:2013wha,
Fujii:2013sra,Fujii:2015bua,Fujii:2015vha,
Alexandru:2015xva,Alexandru:2015sua,Alexandru:2017lqr}. 
The idea is to continuously deform the original integration surface $\bbR^N$ 
(or a compact Lie group $G$) into a surface $\Sigma$ 
within the complexified space $\bbC^N$ (or $G^\bbC$) 
so that the oscillatory behavior of the integrand is significantly mitigated there. 
Although Cauchy's theorem guarantees that 
the value of the integral remains unchanged 
between the original and deformed surfaces, 
Monte Carlo sampling on $\Sigma$ often yields incorrect estimates 
due to ergodicity issues. 
This occurs 
because the deformed surface becomes partitioned by the zeros of the Boltzmann weight 
as the surface is pushed deeper into the complexified space 
to adequately suppress the oscillations. 

A solution to this tradeoff%
---balancing the reduction of the sign problem 
against the emergence of the ergodicity problem---%
was first proposed as the \emph{tempered Lefschetz thimble} (TLT) method 
\cite{Fukuma:2017fjq,Alexandru:2017oyw} 
(see also Ref.~\cite{Fukuma:2019wbv}), 
where the configuration space is extended 
by treating the deformation parameter (called the flow time) 
as an additional dynamical variable. 
In this approach, a finite number of replicas 
(and thus a discrete set of deformation levels) 
are introduced. 
However, a major drawback of the TLT method is its high computational cost: 
to account for the difference in volume elements between replicas, 
one needs to compute the Jacobian of the deformation 
at every configuration exchange between adjacent replicas, 
a process requiring $O(N^3)$ operations. 
Furthermore, maintaining a reasonable exchange acceptance rate 
requires a large number of replicas. 

The Worldvolume Hybrid Monte Carlo (WV-HMC) method 
\cite{Fukuma:2020fez} 
(see also Refs.~\cite{Fukuma:2021aoo,Fukuma:2023eru,Fukuma:2025gya,Fukuma:2025uzg,
Namekawa:2024ert,Fukuma:2025esu}) 
was subsequently proposed 
to overcome these computational bottlenecks. 
In this approach, an $(N+1)$-dimensional submanifold $\calR$ 
is introduced within the complexified space ($\bbC^N$ or $G^\bbC$) 
as a continuous union of deformed surfaces, 
termed the worldvolume. 
Molecular dynamics (MD) is then performed on its tangent bundle $T\calR$, 
which naturally possesses a symplectic structure. 
By employing a symplectic integrator 
(such as RATTLE \cite{Andersen:1983,Leimkuhler:1994}) 
to precisely preserve the symplectic volume form during the MD steps, 
this method completely eliminates the need to compute the Jacobian 
when generating configurations 
(see Refs.~\cite{Fukuma:2023eru,Fukuma:2025gya} for detailed arguments). 

Both the TLT method and its continuous extension, WV-HMC, 
have proven to be reliable and versatile algorithms. 
They have been successfully applied to various models, 
such as the $(1+0)$-dimensional Thirring model 
via TLT \cite{Fukuma:2017fjq}, 
the Stephanov model via WV-HMC \cite{Fukuma:2020fez}, 
and the two-dimensional doped Hubbard model 
via both TLT \cite{Fukuma:2019wbv} and WV-HMC \cite{Fukuma:2025uzg}, 
although the simulated system sizes have thus far remained moderately small. 
The WV-HMC algorithm has also been extended 
to cases where the configuration space is a group manifold 
\cite{Fukuma:2025gya}, 
which can be regarded as the most general setting 
for lattice gauge theories. 

In the present paper, 
we focus on cases 
where the maximum flow time can be restricted to a small value. 
A prime example is the application of WV-HMC 
to the two-dimensional doped Hubbard model 
using a redundant, nonphysical parameter $\alpha$, 
as introduced in Ref.~\cite{Beyl:2017kwp}.%
\footnote{ 
  See Refs.~\cite{Mukherjee:2014hsa,Ulybyshev:2017hbs,Ulybyshev:2019hfm,
  Ulybyshev:2019fte,Ulybyshev:2022kxq,Ulybyshev:2024kdr} 
  for studies of the doped Hubbard model 
  with Lefschetz thimble-based approaches. 
} 
An optimal choice of $\alpha$ 
significantly mitigates the sign problem on the original integration surface 
while preserving ergodicity both on and near this surface,
thereby allowing the maximum flow time to remain small \cite{Fukuma:2025uzg}. 
However, when the worldvolume is compressed into a thin layer, 
phase-space exploration becomes inefficient 
because the momentum refresh in WV-HMC is performed isotropically. 
This inefficiency can cause ergodicity problems to reappear.

To improve ergodicity within a thin worldvolume, 
we propose embedding generalized thimble HMC (GT-HMC) 
\cite{Alexandru:2019,Fukuma:2019uot}
into the WV-HMC framework. 
GT-HMC performs HMC updates on a deformed surface at a fixed flow time 
and thus typically suffers from its own ergodicity issues. 
Nevertheless, it enables highly efficient exploration within the allowed regions, 
and its MD step size can usually be chosen to be much larger than that of WV-HMC. 
Consequently, GT-HMC is well suited for sampling regions 
where ergodicity issues are less severe. 
We provide a rigorous proof 
that GT-HMC can be seamlessly embedded within the WV-HMC algorithm. 
The crux of our argument relies on showing that 
the tangent bundle over a deformed surface (utilized in GT-HMC) 
can be embedded into that over the worldvolume (utilized in WV-HMC)
while strictly preserving the underlying symplectic structure. 
We confirm that the standalone (``pure'') and combined algorithms 
agree within statistical errors 
for the two-dimensional doped Hubbard model on an $8 \times 8$ lattice. 
This combined algorithm enables simulations on larger spacetime lattices. 
We demonstrate the feasibility of this approach 
by extrapolating the number and energy densities
to the zero Trotter step limit 
at fixed temperature $T/t = 1/6.4\simeq 0.156$ 
and repulsive interaction $U/t = 8.0$, 
where $t$ denotes the hopping amplitude. 
Even with modest sample sizes, 
we achieve controlled statistical errors 
across the entire range of the chemical potential.

This paper is organized as follows. 
In Sect.~\ref{sec:gt}, 
we review the Lefschetz thimble method and its generalization%
---the generalized thimble method \cite{Alexandru:2015sua}. 
In Sect.~\ref{sec:embedding}, 
after giving a brief overview of GT-HMC 
\cite{Alexandru:2019,Fukuma:2019uot} 
and WV-HMC \cite{Fukuma:2020fez}, 
we prove that GT-HMC can be embedded into WV-HMC symplectically. 
In Sect.~\ref{sec:application}, 
we review the quantum Monte Carlo computations 
of the two-dimensional doped Hubbard model, 
following Ref.~\cite{Fukuma:2025uzg}. 
In Sect.~\ref{sec:numerical_tests}, 
we apply the combined algorithm to the model 
on an $8 \times 8$ lattice 
at $T/t = 1/6.4\simeq 0.156$ and $U/t = 8.0$. 
Section~\ref{sec:conclusions} is devoted to conclusions and outlook.

\section{Generalized thimble method}
\label{sec:gt}

In this section, 
we begin with a brief explanation of the sign problem 
and then introduce the Lefschetz thimble method 
\cite{Witten:2010cx,Cristoforetti:2012su,Cristoforetti:2013wha,
Fujii:2013sra,Fujii:2015bua,Fujii:2015vha,Alexandru:2015xva}
as well as its generalization---the generalized thimble method 
\cite{Alexandru:2015sua}.

\subsection{Lefschetz thimble method and its generalization}
\label{sec:lt}

Our aim is to numerically evaluate 
the expectation value of an observable $\calO(x)$ 
defined by a path integral 
over the configuration space $\mathbb{R}^N = \{x = (x^a)\}$:%
\footnote{ 
  We omit the symbol $\wedge$ (wedge) 
  when the omission does not cause confusion. 
} 
\begin{align}
  \langle \calO \rangle
  \equiv \frac{\int_{\bbR^N} dx \, e^{-S(x)} \, \calO(x)}
  {\int_{\bbR^N} dx \, e^{-S(x)}}
  \quad
  \Bigl(dx = dx^1 \wedge \cdots \wedge dx^N \equiv \prod_{a=1}^N dx^a \Bigr),
\label{vev}
\end{align}%
where $S(x) \in \bbC$ is a complex-valued action.
The Boltzmann weight $e^{-S(x)} / \int dx\,e^{-S(x)}$ 
does not define a real and positive probability measure, 
which prevents the direct application of Markov chain Monte Carlo methods. 
A standard workaround is the so-called naive reweighting method, 
in which the real part of the action, $\ReS(x)$, 
is used to define the probability measure. 
The expectation value $\vev{\calO}$ is then expressed 
as a ratio of reweighted averages,
\begin{align}
  \vev{\calO} 
  = \frac{\vev{e^{-i\,\ImS(x)}\,\calO(x)}_\mathrm{rewt}}
    {\vev{e^{-i\,\ImS(x)}}_\mathrm{rewt}},
\label{rewt_naive1}
\end{align}%
where $\vev{\cdots}_\mathrm{rewt}$ is defined by 
\begin{align}
  \vev{g(x)}_\mathrm{rewt} = \frac{\int_{\bbR^N} dx\, e^{-\ReS(x)}\,g(x)}
  {\int_{\bbR^N} dx\,e^{-\ReS(x)}}.
\label{rewt_naive2}
\end{align}%
For systems with a large number of degrees of freedom ($N \gg 1$), 
both the numerator and the denominator in Eq.~\eqref{rewt_naive1} 
involve highly oscillatory integrals, 
yielding exponentially small values of order $e^{-O(N)}$. 
When they are evaluated by Monte Carlo sampling, 
the resulting statistical errors are of order $O(1/\sqrt{N_\mathrm{conf}})$,
where $N_\mathrm{conf}$ is the sample size. 
Thus, to achieve a reasonable signal-to-noise ratio, 
an exponentially large number of configurations  
($N_\mathrm{conf} \gtrsim e^{O(N)}$) 
is required. 
This is the sign problem considered in this paper. 

\begin{figure}[t]
  \centering
  \includegraphics[width=90mm]{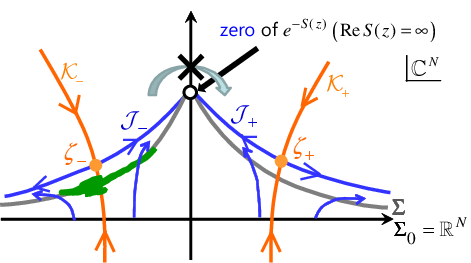}
  \caption{%
    Lefschetz thimble method and its ergodicity problem. 
    The original surface $\Sigma_0 = \bbR^N$ 
    is deformed into $\Sigma$ within $\bbC^N$.
    $\calJ_{\pm}$ ($\calK_{\pm}$) are 
    Lefschetz thimbles (anti-Lefschetz thimbles) 
    associated with critical points $\zeta_\pm$. 
    A Monte Carlo walker cannot move from the vicinity of one thimble $\calJ_-$
    to that of another thimble $\calJ_+$ 
    due to the infinitely high potential barrier at the zeros of $e^{-S(z)}$ 
    (figure adapted from \cite{Fukuma:2023eru}).
  }
  \label{fig:ergodicity_problem}
\end{figure}

In the Lefschetz thimble method, 
the integration surface $\Sigma_0 = \bbR^N$ 
is continuously deformed into a new surface $\Sigma$ 
within the complexified space $\bbC^N$ 
so that the oscillatory behavior of the integrand 
is reduced there (see Fig.~\ref{fig:ergodicity_problem}). 
Provided that 
both $e^{-S(z)}$ and $e^{-S(z)}\,\calO(z)$ are holomorphic functions on $\bbC^N$ 
(which is usually the case for physically interesting models), 
Cauchy's theorem guarantees that 
the integrals remain invariant under the deformation: 
\begin{align}
  \vev{\calO}
  = \frac{\int_{\Sigma} dz \, e^{-S(z)} \, \calO(z)}
  {\int_{\Sigma} dz \, e^{-S(z)}}
  \quad
  (dz = dz^1 \wedge \cdots \wedge dz^N).
\label{cauchy}
\end{align}%
The sign problem is thus expected to be significantly alleviated 
if the imaginary part $\ImS(z)$ is nearly constant 
on the deformed surface $\Sigma$. 
Such a deformation is realized 
by the anti-holomorphic gradient flow:  
\begin{align}
  \dot{z} &= \overline{\partial S(z)}
  ~~\mbox{with}~~
  z|_{t=0} = x,
\label{flow_config}
\end{align}%
where $\dot{z} = \partial z/\partial t$ 
($t$: the deformation parameter referred to as the \emph{flow time}), 
and $x$ denotes the initial configuration on $\bbR^N$. 
A point on the flow is uniquely specified as $z = z(t,x)$. 
This flow equation satisfies the (in)equality 
\begin{align}
  [S(z)]^\centerdot = \partial S(z)\cdot \dot{z}
  = |\partial S(z)|^2 \geq 0,
\end{align}%
which implies that 
the real part $\ReS(z)$ always increases along the flow 
except at critical points 
[where the gradient $\partial S(z)$ vanishes], 
while the imaginary part $\ImS(z)$ remains constant along the flow. 
We define the Lefschetz thimble $\calJ$ associated with a critical point $\zeta$ 
as the set of points that flow out from $\zeta$ 
(see Fig.~\ref{fig:ergodicity_problem}), 
on which $\ImS(z)$ is constant due to its invariance under the flow. 
As the flow time $t$ increases, 
the deformed surface $\Sigma_t \equiv \{ z(t,x) \mid x \in \Sigma_0 \}$ 
approaches a union of Lefschetz thimbles. 
Thus, the oscillatory behavior of the integrand 
is expected to be significantly reduced 
if the flow is evolved for a sufficiently large time. 

In the original Lefschetz thimble method 
\cite{Cristoforetti:2012su,Cristoforetti:2013wha,
Fujii:2013sra,Fujii:2015bua,Fujii:2015vha}, 
sampling is performed directly on a Lefschetz thimble. 
However, $\ReS(z)$ diverges at the boundaries between adjacent Lefschetz thimbles, 
which leads to a serious ergodicity problem. 
The \emph{generalized thimble method} \cite{Alexandru:2015sua} 
proposes to choose a deformed surface at an intermediate flow time 
that is large enough to suppress the sign problem, 
but not so large that ergodicity issues arise. 
A closer study \cite{Fukuma:2019wbv} shows, however, that 
the sign problem is not significantly mitigated 
until the deformed surface touches the zeros of the Boltzmann weight 
(including points at infinity), 
which makes it extremely difficult to find such an ideal flow time 
in practical applications (such as the Hubbard model). 
Nevertheless, the generalized thimble method retains its value, 
because it allows one to estimate the degree of sign-problem reduction 
by monitoring the average phase factor as a function of the flow time. 
Furthermore, as we argue in subsequent sections, 
the generalized thimble algorithm can be exploited 
to \emph{enhance} the ergodicity of WV-HMC 
by embedding it into the worldvolume algorithm 
(\emph{similia similibus curantur}). 

\begin{figure}[t]
  \centering
  \includegraphics[width=70mm]{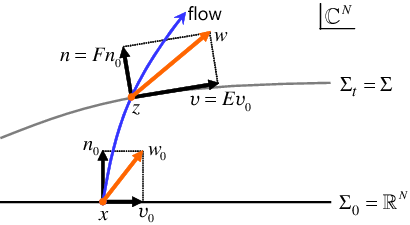}
  \caption{%
    Deformation of the integration surface 
    and the lift of tangent and normal vectors  
    (figure adapted from Ref.~\cite{Fukuma:2023eru}).
  }
\label{fig:multEF}
\end{figure}

In sampling $\Sigma = \Sigma_t$, 
we need to lift 
a tangent vector $v_0 = (v_0^a) \in T_x \Sigma_0$ 
and a normal vector $n_0 =(n_0^a) \in N_x \Sigma_0$
to 
$v = E v_0 \in T_z \Sigma$ ($v^i = E^i_a v_0^a$) 
and $n = F n_0 \in N_z \Sigma$ ($n^i = F^i_a n_0^a$),
respectively 
(see Fig.~\ref{fig:multEF}). 
%
This lifting can be realized by the following flow equations 
\cite{Cristoforetti:2012su,Fujii:2013sra,Fukuma:2023eru}: 
\begin{align}
  \dot{v} &= \overline{H(z) v}
  ~~\mbox{with}~~
  v|_{t=0} = v_0,
\label{flow_tangent}
\\
  \dot{n} &= -\overline{H(z) n}
  ~~\mbox{with}~~
  n|_{t=0} = n_0,
\label{flow_normal}
\end{align}%
where $H_{ij}(z)\equiv \partial_i \partial_j S(z)$ is the Hessian matrix.
In the flat case, 
the matrices $E = (E^i_a)$ and $F = (F^i_a)$ 
are given by the Jacobian matrix: 
\begin{align}
  E_a^i = F_a^i = \frac{\partial z^i}{\partial x^a},
\end{align}%
which can be evaluated 
by integrating the following flow equation 
for the vectors $E_a = (E_a^i)$:
\begin{align}
  \dot{E}_a = \overline{H(z) E_a}
  \with
  E_a^i \bigr|_{t=0} = \delta^i_a.
\end{align}%
%

\subsection{More on the generalized thimble method}
\label{sec:gt_more}

In the generalized thimble method \cite{Alexandru:2015sua}, 
we consider a deformed surface $\Sigma = \Sigma_t$ at a fixed flow time $t$. 
In the following discussion, 
we denote $z = z(t,x)$ simply by $z = z(x)$, 
unless the dependence on $t$ needs to be made explicit. 

We rewrite the expression \eqref{cauchy}
as a ratio of reweighted averages on the deformed surface $\Sigma$: 
\begin{align}
  \vev{\calO} 
  = \frac{\vev{ \calF_\Sigma(z)\,\calO(z) }_\Sigma}
  {\vev{ \calF_\Sigma(z) }_\Sigma},
\label{rewt_Sigma1}
\end{align}%
where $\vev{\cdots}_\Sigma$ is defined by 
\begin{align}
  \vev{g(z)}_\Sigma 
  = \frac{\int_\Sigma |dz|_\Sigma \, e^{-\ReS(z)}\,g(z)}
  {\int_\Sigma |dz|_\Sigma \,e^{-\ReS(z)}}.
\label{vev_Sigma1}
\end{align}%
Here, $|dz|_\Sigma$ is the invariant measure on $\Sigma$, 
and $\calF_\Sigma(z) \equiv dz\,e^{-i\,\ImS(z)} / |dz|_\Sigma$ 
is the associated reweighting factor.  

The invariant measure $|dz|_\Sigma$ is defined as follows. 
We first introduce the inner product between complex vectors 
$u = (u^i)$ and $v = (v^i) \in \bbC^N$ as 
\begin{align}
  \vev{u,v} \equiv \re u^\dagger v = \re\overline{u^i} v^i,
\end{align}%
with which the flat metric on the complex space $\bbC^N$ 
is expressed as 
$ds^2 = \vev{dz,dz} = \overline{dz^i} dz^i$. 
The induced metric associated with the embedding $z = z(x)$ is then given by%
\footnote{ 
  The coordinates $x = (x^a)$ on $\Sigma$ can be chosen arbitrarily 
  as long as they provide a unique parametrization of $\Sigma$. 
  In Ref.~\cite{Alexandru:2015sua}, 
  the chosen coordinates were vectors in the tangent space 
  at the critical point of the dominant Lefschetz thimble, 
  and the corresponding algorithm was called the \emph{contraction algorithm}.
  The approach of using initial configurations as coordinates 
  was first systematically employed in Ref.~\cite{Fukuma:2017fjq}. 
} 
\begin{align}
  ds_\Sigma^2 \equiv \vev{dz(x),dz(x)} \equiv \gamma_{ab}(x) dx^a dx^b,
\end{align}%
where $\gamma_{ab}$ is expressed 
in terms of the complex tangent vectors $E_a = (E_a^i)$ as 
\begin{align}
  \gamma_{ab} = \vev{E_a, E_b}. 
\end{align}%
The invariant measure is then given by
\begin{align}
  |dz|_\Sigma = dx\,\sqrt{\gamma(x)}
  \quad
  (\gamma = \det(\gamma_{ab})).
\label{measure_Sigma}
\end{align}%
The reweighting factor $\calF_\Sigma(z)$ thus takes the form 
\begin{align}
  \calF_\Sigma(z) = \frac{dz}{|dz|_\Sigma}\,e^{-i\,\ImS(z)}
  = \frac{\det E}{\sqrt{\gamma}}\,e^{-i\,\ImS(z)}.
\end{align}%
In the flat case $\Sigma_0 = \bbR^N$, 
we have $\sqrt{\gamma} = |\det E|$ \cite{Fujii:2013sra} 
(see also Ref.~\cite{Fukuma:2023eru}), 
and thus the reweighting factor is a pure phase factor:
\begin{align}
  \calF_\Sigma(z) = \frac{\det E}{|\det E|}\,e^{-i\,\ImS(z)}.
\label{rewtfactor_GT}
\end{align}%

\section{Embedding GT-HMC into the WV-HMC framework}
\label{sec:embedding}

In this section, 
we first briefly review the generalized thimble Hybrid Monte Carlo (GT-HMC) method 
and the Worldvolume Hybrid Monte Carlo (WV-HMC) method, 
and then show that GT-HMC can be embedded into the WV-HMC framework.

\subsection{Generalized thimble Hybrid Monte Carlo (GT-HMC)}
\label{sec:gt-hmc}

In this subsection, 
we present a brief overview of GT-HMC 
\cite{Alexandru:2019,Fukuma:2019uot}. 
For more detailed explanations 
presented in a style similar to that of this paper, 
see Refs.~\cite{Fukuma:2023eru,Fukuma:2025gya}.

The GT-HMC method 
\cite{Alexandru:2019,Fukuma:2019uot,Fukuma:2023eru,Fukuma:2025gya} 
is an HMC algorithm 
that performs sampling on a deformed surface $\Sigma = \Sigma_t$. 
This method generalizes the original Lefschetz thimble HMC method 
\cite{Fujii:2013sra}, 
which performs sampling directly on a single dominant Lefschetz thimble. 
To define Hamiltonian dynamics,
we introduce a momentum $p = (p_a)$ $(a=1,\ldots,N)$ 
and rewrite the measure \eqref{measure_Sigma} in the form%
\begin{align}
  |dz|_\Sigma \propto dx\, dp\, e^{-(1/2)\, \gamma^{ab} p_a p_b}
  = d\Omega_\Sigma\,e^{-(1/2)\, \gamma^{ab} p_a p_b}, 
\end{align}%
where $dp \equiv dp_1 \wedge \cdots \wedge dp_N \equiv \prod_a dp_a$, 
$(\gamma^{ab}) \equiv (\gamma_{ab})^{-1}$, 
and $d\Omega_\Sigma$ is the symplectic volume form defined by 
\begin{align}
  d\Omega_\Sigma \equiv \frac{\omega_\Sigma^N}{N!}
  ~(= dx\,dp)
\label{dOmega_Sigma}
\end{align}%
with the symplectic 2-form 
\begin{align}
  \omega_\Sigma \equiv d(p_a dx^a) = dp_a \wedge dx^a.
\end{align}%
The reweighted average on $\Sigma$ [Eq.~\eqref{vev_Sigma1}] 
can then be written as a phase-space integral: 
\begin{align}
  \vev{g(z)}_\Sigma 
  = \frac{\int d\Omega_\Sigma\,e^{-H_0^\mathrm{GT}(x,p)}\,g(z(x))}
    {\int d\Omega_\Sigma\,e^{-H_0^\mathrm{GT}(x,p)}},
\label{vev_Sigma2}
\end{align}%
where the Hamiltonian is given by 
\begin{align}
  H_0^\mathrm{GT}(x,p) = \frac{1}{2}\, \gamma^{ab}(x)\,p_a p_b + \ReS(z(x)).
\end{align}%

The expression \eqref{vev_Sigma2} can be further rewritten 
as an integral over the tangent bundle of $\Sigma$, 
\begin{align}
  T\Sigma = \{(z,\pi) \mid z\in \Sigma,\,\pi\in T_z\Sigma\}.
\end{align}%
To this end, 
we lift the vector 
$p^a \equiv \gamma^{ab} p_b$ 
to $\pi = (\pi^i) \in T_z \Sigma$ $(i=1,\ldots,N)$ as 
\begin{align}
  \pi^i = E_a^i\,p^a.
\end{align}%
One can show that 
the symplectic potential $a_\Sigma = p_a dx^a$ can be rewritten as 
\begin{align}
  a_\Sigma = \vev{\pi, dz}, 
\end{align}%
and thus the symplectic form $\omega_\Sigma = da_\Sigma$ becomes 
\begin{align}
  \omega_\Sigma = d \vev{\pi, dz} = \re( \overline{d\pi^i} \wedge dz^i ). 
\end{align}%
Furthermore, the kinetic term $(1/2)\, \gamma^{ab}\,p_a p_b$ 
can be expressed as $(1/2)\, \vev{\pi,\pi}$. 
Thus, the reweighted average becomes 
\begin{align}
  \vev{g(z)}_\Sigma 
  = \frac{\int_{T\Sigma} d\Omega_\Sigma\,e^{-H^\mathrm{GT}(z,\pi)}\,g(z)}
  {\int_{T\Sigma} d\Omega_\Sigma\,e^{-H^\mathrm{GT}(z,\pi)}}
\label{vev_Sigma3}
\end{align}%
with the Hamiltonian now given by 
\begin{align}
  H^\mathrm{GT}(z,\pi) = \frac{1}{2}\,\vev{\pi,\pi} + \ReS(z). 
\end{align}%

The GT-HMC method 
\cite{Alexandru:2019,Fukuma:2019uot,Fukuma:2023eru,Fukuma:2025gya} 
can be viewed 
as defining a stochastic process on $T\Sigma$ 
whose equilibrium distribution is given by%
\footnote{ 
  See Ref.~\cite{Fukuma:2023eru}, especially Ref.~\cite{Fukuma:2025gya}, 
  for a detailed discussion of this viewpoint. 
} 
\begin{align}
 \rho^\mathrm{GT}(z,\pi) 
 = \frac{ e^{-H^\mathrm{GT}(z,\pi)} }
 { \int_{T\Sigma}\,d\Omega_\Sigma\,e^{-H^\mathrm{GT}(z,\pi)} },
\end{align}%
where the Hamiltonian has the time-reversal symmetry 
$H^\mathrm{GT}(z,-\pi) = H^\mathrm{GT}(z,\pi)$. 
The transition probability $P^\mathrm{GT}(z',\pi' \mid z,\pi)$ 
is required to satisfy a suitable ergodicity condition 
that ensures convergence to the unique equilibrium distribution. 
A practical way to construct such a stochastic process 
is to decompose it into a sequence of subprocesses 
$P^\mathrm{GT}_{(k)}(z',\pi' \mid z,\pi)$ $(k=1,\ldots,K)$ 
such that 
each satisfies the detailed balance condition for MD, 
\begin{align}
  P^\mathrm{GT}_{(k)}(z',\pi' \mid z,\pi)\,e^{-H^\mathrm{GT}(z,\pi)}
  =  P^\mathrm{GT}_{(k)}(z,-\pi \mid z',-\pi')\,e^{-H^\mathrm{GT}(z',-\pi')},
\label{db_gt}
\end{align}%
and that their composition 
$P^\mathrm{GT} \equiv P^\mathrm{GT}_{(K)}\cdots P^\mathrm{GT}_{(1)}$
has the required ergodicity property. 
We choose the following two stochastic processes as such subprocesses:   
\begin{itemize}
\item
  \underline{Heat bath for $\pi$}:%
  \footnote{ 
    $\delta_\Sigma(z'-z)$ is proportional to $\delta(x'-x)$. 
    Jacobian factors can be neglected in the argument for detailed balance  
    because $z' = z$ in Eq.~\eqref{db_gt} \cite{Fukuma:2025gya}. 
  } 
  \begin{align} \vspace{-2ex}
    P^\mathrm{GT}_{(1)}(z',\pi' \mid z,\pi) 
    = e^{-(1/2)\, \vev{\pi',\pi'}}\,\delta_\Sigma(z'-z),
  \label{gt_hb}
  \end{align}%

\item
  \underline{MD followed by Metropolis test}:%
  \footnote{ 
    The transition probability for the case $(z',\pi') = (z,\pi)$ 
    is determined by the normalization condition 
    $\int_{T\Sigma} d\Omega'\,P^\text{GT}_{(2)} (z',\pi'\,|\,z,\pi) = 1$. 
  } 
  \begin{align}
    P^\mathrm{GT}_{(2)}(z',\pi' \mid z,\pi) 
    &= \min\bigl(1,e^{-H^{\text{GT}}(z',\pi')+H^{\text{GT}}(z,\pi)}\bigr)\,
      \delta_{T\Sigma}((z',\pi') - f_\Sigma(z,\pi))
  \nonumber
  \\
    &\hspace{30mm}\mbox{for}~~
    (z',\pi') \neq (z,\pi),
  \label{gt_md}
  \end{align}%
  where $\delta_{T\Sigma}(z,\pi)$ is the symplectic delta function 
  associated with the symplectic volume form $d\Omega_\Sigma$. 
  $f_{\Sigma}$ denotes the MD integrator, 
  which is assumed to be both reversible and volume-preserving.
\end{itemize}
The stochastic processes $P^\mathrm{GT}_{(1)}$ and $P^\mathrm{GT}_{(2)}$ 
satisfy the detailed balance condition \eqref{db_gt}, 
as shown in Ref.~\cite{Fukuma:2025gya} for more general cases. 

\begin{figure}[ht]
  \centering
  \includegraphics[width=70mm]{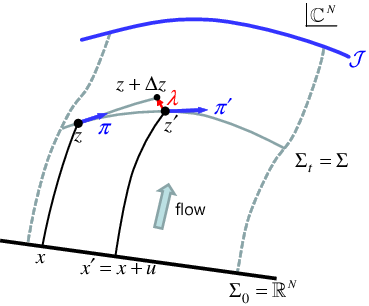}
  \caption{%
  	RATTLE in GT-HMC 
    (figure adapted from Ref.~\cite{Fukuma:2023eru}).
  }
  \label{fig:rattle_gt_wo_tpi}
\end{figure}

MD on $T\Sigma$ of step size $\Delta s$ 
is implemented by repeating the RATTLE update 
\cite{Andersen:1983,Leimkuhler:1994} of the following form 
(see Fig.~\ref{fig:rattle_gt_wo_tpi})
\cite{Alexandru:2019,Fukuma:2019uot,Fukuma:2023eru}: 
\begin{align}
  \pi_{1/2} 
  &= \pi 
  - \frac{\Delta s}{2}\,\overline{\partial S(z)} - \Delta s\,\lambda,
  \label{gt_rattle1}
  \\
  z' &= z + \Delta s\,\pi_{1/2},
  \label{gt_rattle2}
  \\
  \pi' 
  &= \pi_{1/2} 
  - \frac{\Delta s}{2}\,\overline{\partial S(z')} - \Delta s\,\lambda',
  \label{gt_rattle3}
\end{align}%
where the Lagrange multipliers $\lambda$ and $\lambda'$ 
are determined 
so that $z' \in \Sigma$ and $\pi' \in T_{z'}\Sigma$, respectively 
(see Refs.~\cite{Fukuma:2019uot,Fukuma:2023eru} 
for explicit algorithms).  
This RATTLE update is 
(i) exactly reversible, 
(ii) symplectic (and thus volume-preserving), 
and 
(iii) energy-conserving up to second order in $\Delta s$, 
$H^\mathrm{GT}(z',\pi') - H^\mathrm{GT}(z,\pi) = O(\Delta s^3)$.
%

\subsection{Worldvolume Hybrid Monte Carlo (WV-HMC)}
\label{sec:wv-hmc}

In this subsection, 
we present a brief overview of WV-HMC 
\cite{Fukuma:2020fez,Fukuma:2021aoo,Fukuma:2023eru,Fukuma:2025gya}. 
For a more detailed exposition presented in a style aligned with the present paper, 
see Refs.~\cite{Fukuma:2023eru,Fukuma:2025gya}.

\begin{figure}[t]
  \centering
  \includegraphics[width=90mm]{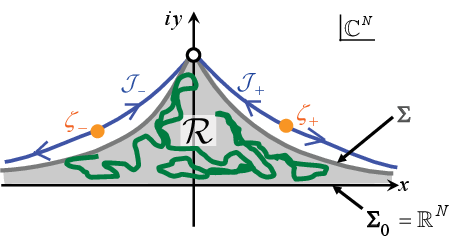}
  \caption{%
   Worldvolume $\calR$
   (figure adapted from Ref.~\cite{Fukuma:2023eru}).
  }
  \label{fig:worldvolume}
\end{figure}

We start from the following expression [Eq.~\eqref{cauchy}]:
\begin{align}
  \vev{\calO}
  = \frac{\int_{\Sigma_t} dz \, e^{-S(z)} \, \calO(z)}
  {\int_{\Sigma_t} dz \, e^{-S(z)}}
  \quad
  (dz = dz^1 \wedge \cdots \wedge dz^N).
\label{cauchy2}
\end{align}%
Here, $\Sigma_t$ denotes the deformed surface at flow time $t$. 
Since both the numerator and the denominator 
in Eq.~\eqref{cauchy2} are independent of $t$,
we can average over $t$ 
with an arbitrary common weight $e^{-W(t)}$ 
\cite{Fukuma:2020fez}:
\begin{align}
  \vev{\calO}
  = \frac{\int dt \, e^{-W(t)} \int_{\Sigma_t} dz \, e^{-S(z)}\,\calO(z)}
  {\int dt \, e^{-W(t)} \int_{\Sigma_t} dz \, e^{-S(z)}} .
\label{observable_WV_HMC}
\end{align}%
This expression can be written as a ratio of 
reweighted averages on the \emph{worldvolume} $\calR$ 
(see Fig.~\ref{fig:worldvolume}), 
\begin{align}
  \calR\equiv \bigcup_{t} \Sigma_t = \{z(t,x) \mid t\in\bbR,\,x\in\bbR^N\}, 
\end{align}%
and becomes 
\begin{align}
  \vev{\calO} 
  = \frac{\vev{ \calF_\calR(z)\,\calO(z) }_\calR}
  {\vev{ \calF_\calR(z) }_\calR},
\label{rewt_calR}
\end{align}%
where $\vev{\cdots}_\calR$ is defined by 
\begin{align}
  \vev{g(z)}_\calR
  = \frac{\int_\calR |dz|_\calR \, e^{-\ReS(z) - W(t)}\,g(z)}
  {\int_\calR |dz|_\calR \,e^{-\ReS(z) - W(t)}}.
\label{vev_calR1}
\end{align}%
Here, $|dz|_\calR$ is the invariant measure on $\calR$, 
and $\calF_\calR(z) \equiv dt\,dz\,e^{-i\,\ImS(z)} / |dz|_\calR$ 
is the associated reweighting factor. 

The invariant measure $|dz|_\calR$ is given as follows. 
We first introduce coordinates on $\calR$ as 
$\hat{x} = (\hat{x}^\mu) = (\hat{x}^0 = t, \hat{x}^a = x^a) = (t,x)$. 
The induced metric on $\calR$ 
associated with the embedding $z = z(\hat{x})$ 
is then given by
\begin{align}
  ds_\calR^2 
  \equiv \vev{dz(\hat{x}),dz(\hat{x})} 
  \equiv \hat\gamma_{\mu\nu}(\hat{x}) d\hat{x}^\mu d\hat{x}^\nu,
\label{metric_calR}
\end{align}%
where $\hat{\gamma}_{\mu\nu}$ is expressed 
in terms of the complex vectors
\begin{align}
  \hat{E}_\mu = (\hat{E}_\mu^i \equiv \partial z^i / \partial \hat{x}^\mu) 
\end{align}%
as 
\begin{align}
  \hat\gamma_{\mu\nu} = \vev{\hat{E}_\mu, \hat{E}_\nu}.
\end{align}%
The invariant measure is then given by 
\begin{align}
  |dz|_\calR = d\hat{x}\,\sqrt{\hat\gamma}
  \quad
  (\hat\gamma = \det(\hat\gamma_{\mu\nu})),
\label{measure_calR1}
\end{align}%
where $d\hat x = d\hat x^0 \wedge d\hat x^1 \wedge \cdots \wedge d\hat x^N
\equiv \prod_{\mu=0}^N d\hat x^\mu = dt\,dx$.

The reweighting factor $\calF_\calR(z)$ 
is expressed as follows. 
By using the flow equation $\dot{z}^i = \xi^i(z)$ 
with the flow vector 
\begin{align}
  \xi^i(z) \equiv \overline{\partial_i S(z)}, 
\end{align} 
the complex vector 
$\hat{E}_\mu = (\hat{E}_\mu^i = \partial z^i / \partial \hat{x}^\mu)$ 
can be written as 
\begin{align}
  \hat{E}_0 = \xi,
  \quad
  \hat{E}_a = E_a.
\end{align}%
We decompose $\xi$ into its tangential and normal components as 
\begin{align}
  \xi = \xi_v + \xi_n
  \quad
  (\xi_v \in T_z \Sigma,\,\,\xi_n \in N_z \Sigma),
\end{align}%
from which we obtain 
\begin{align}
  dz^i = \xi^i dt + E_a^i dx^a
  = \xi_n^i dt + E_a^i (dx^a + \beta^a dt) 
\label{dz_decomposition}
\end{align}%
with $\beta^a \equiv \gamma^{ab} \vev{E_b,\xi_v}$. 
Then, the metric \eqref{metric_calR} takes the ADM form: 
\begin{align}
  ds_\calR^2 
  = \alpha^2 dt^2 + \gamma_{ab}\,(dx^a + \beta^a dt) (dx^b + \beta^b dt),
\end{align}%
where the induced metric $\gamma_{ab}$ on $\Sigma$, 
the shift vector $\beta^a$, 
and the lapse function $\alpha$ 
are given by 
\begin{align}
  \gamma_{ab} = \vev{E_a,E_b},
  \quad
  \beta^a = \gamma^{ab}\,\vev{E_b,\xi_v},
  \quad
  \alpha^2 = \vev{\xi_n,\xi_n}. 
\end{align}%
The invariant measure \eqref{measure_calR1} can then be rewritten as 
\begin{align}
  |dz|_\calR = \alpha \sqrt{\gamma}\,dt\,dx. 
\label{measure_calR2}
\end{align}%
The reweighting factor $\calF_\calR(z)$ in Eq.~\eqref{rewt_calR} 
is thus expressed as%
\footnote{ 
  Recall that $\sqrt{\gamma} = |\det E|$ 
  for the flat case $\Sigma_0 = \bbR^N$, 
  so that $dz = \det E\cdot dx = (\det E / |\det E|)\,\sqrt{\gamma}\, dx$. 
} 
\begin{align}
  \calF_\calR(z) 
  = \frac{dt\, dz}{|dz|_\calR}\,e^{-i\,\ImS(z)}
  = \alpha^{-1}\,\frac{\det E}{|\det E|}\,e^{-i\,\ImS(z)}.
\label{rewtfactor_R}
\end{align}%

The WV-HMC method 
\cite{Fukuma:2020fez}
is an algorithm 
that performs HMC sampling on the worldvolume $\calR$. 
To define the Hamiltonian dynamics,
we introduce a momentum $\hat{p} = (\hat{p}_\mu)$ $(\mu=0,1,\ldots,N)$ 
and rewrite the measure \eqref{measure_calR1} as 
\begin{align}
  |dz|_\calR 
  \propto d\hat{x}\, d\hat{p}\,
  e^{-(1/2)\, \hat\gamma^{\mu\nu} \hat{p}_\mu \hat{p}_\nu}
  = d\Omega_\calR\,e^{-(1/2)\,  \hat\gamma^{\mu\nu} \hat{p}_\mu \hat{p}_\nu}. 
\end{align}%
Here, $d\hat{p} \equiv d\hat{p}_0 \wedge d\hat{p}_1 \wedge 
\cdots \wedge d\hat{p}_N \equiv \prod_\mu d\hat{p}_\mu$, 
$(\hat\gamma^{\mu\nu}) \equiv (\hat\gamma_{\mu\nu})^{-1}$, 
and $d\Omega_\calR$ is the symplectic volume form defined by 
\begin{align}
  d\Omega_\calR \equiv \frac{\omega_\calR^{N+1}}{(N+1)!}
  ~ ( = d\hat x\,d\hat p)
\label{dOmega_calR}
\end{align}%
with the symplectic 2-form  
\begin{align}
  \omega_\calR \equiv d(\hat{p}_\mu d\hat{x}^\mu) 
  = d\hat{p}_\mu \wedge d\hat{x}^\mu.
\end{align}%
The reweighted average on $\calR$ [Eq.~\eqref{vev_calR1}] 
is then expressed as a phase-space integral: 
\begin{align}
  \vev{g(z)}_\calR 
  = \frac{\int d\Omega_\calR\,e^{-H_0(\hat{x},\hat{p})}\,g(z(\hat{x}))}
  {\int d\Omega_\calR\,e^{-H_0(\hat{x},\hat{p})}},
\label{vev_calR2}
\end{align}%
where the Hamiltonian is given by 
\begin{align}
  H_0(\hat{x},\hat{p}) 
  = \frac{1}{2}\, \hat\gamma^{\mu\nu} \hat{p}_\mu \hat{p}_\nu 
  + \ReS(z(\hat{x})) + W(t).
\end{align}%

As in GT-HMC, 
the expression \eqref{vev_calR2} can be further rewritten 
as an integral over the tangent bundle of $\calR$
\cite{Fukuma:2020fez,Fukuma:2023eru,Fukuma:2025gya}, 
\begin{align}
  T\calR = \{ (z,\pi) \mid z\in \calR,\, \pi\in T_z\calR \}.
\end{align}%
To this end, 
we define the momentum $\pi = (\pi^i)$ $(i=1,\ldots,N)$ 
on the tangent space $T_z \calR$ by 
\begin{align}
  \pi^i = \hat{E}_\mu^i\,\hat{p}^\mu
  \quad
  (\hat{p}^\mu \equiv \hat\gamma^{\mu\nu}\,\hat{p}_\nu).
\end{align}%
Then, the symplectic potential $a_\calR = \hat{p}_\mu d\hat{x}^\mu$ 
can be rewritten as 
\begin{align}
  a_\calR = \vev{\pi, dz}, 
\end{align}%
and thus the symplectic form $\omega_\calR = da_\calR$ becomes 
\begin{align}
  \omega_\calR 
  = d \vev{\pi, dz} 
  = \re\bigl( \overline{d\pi^i} \wedge dz^i \bigr). 
\end{align}%
Furthermore, the kinetic term 
$(1/2)\, \hat\gamma^{\mu\nu}\,\hat{p}_\mu \hat{p}_\nu$ 
can be expressed as $(1/2)\, \vev{\pi,\pi}$. 
Thus, the reweighted average becomes 
\begin{align}
  \vev{g(z)}_\calR 
  = \frac{\int_{T\calR} d\Omega_\calR\,e^{-H(z,\pi)}\,g(z)}
  {\int_{T\calR} d\Omega_\calR\,e^{-H(z,\pi)}},
\label{vev_calR3}
\end{align}%
where the Hamiltonian is now given by 
\begin{align}
  H(z,\pi) = \frac{1}{2}\,\vev{\pi,\pi} + V(z) 
\end{align}%
with the potential%
\footnote{ 
  Here, $t(z)$ denotes the flow time 
  corresponding to configuration $z = z(t,x)$.
} 
\begin{align}
  V(z) = \ReS(z) + W(t(z)). 
\end{align}%

In parallel with the GT-HMC case, 
the WV-HMC method can be regarded as defining a stochastic process on $T\calR$ 
whose equilibrium distribution is given by%
\footnote{ 
  See Ref.~\cite{Fukuma:2023eru}, especially Ref.~\cite{Fukuma:2025gya}, 
  for a detailed discussion on this viewpoint.
} 
\begin{align}
  \rho(z,\pi) 
   = \frac{ e^{-H(z,\pi)} }{ \int_{T\calR}\,d\Omega_\calR\,e^{-H(z,\pi)} },
\end{align}%
where the Hamiltonian satisfies the time-reversal symmetry, $H(z,-\pi) = H(z,\pi)$. 
The transition probability $P(z',\pi' \mid z,\pi)$ 
is required to satisfy an appropriate ergodicity condition 
that ensures convergence to the unique equilibrium distribution. 
As in GT-HMC, 
such a process can be constructed 
as a composition of subprocesses $P_{(k)}$ 
that satisfy the detailed balance condition for MD: 
\begin{align}
  P_{(k)}(z',\pi' \mid z,\pi)\,e^{-H(z,\pi)}
  =  P_{(k)}(z,-\pi \mid z',-\pi')\,e^{-H(z',-\pi')}. 
\label{db_wv}
\end{align}%
We consider the following two subprocesses: 
\begin{itemize}
  \item
  \underline{Heat bath for $\pi$}:%
  \footnote{ 
    $\delta_\calR(z'-z)$ is proportional to $\delta(t'-t)\,\delta(x'-x)$. 
    Jacobian factors can be neglected in the argument for detailed balance  
    because $z' = z$ in Eq.~\eqref{db_wv} \cite{Fukuma:2025gya}. 
  } 
  \begin{align} \vspace{-2ex}
    P_{(1)}(z',\pi' \mid z,\pi) 
    = e^{-(1/2) \vev{\pi',\pi'}}\,\delta_\calR(z'-z),
  \label{wv_hb}
  \end{align}%
  
  \item
  \underline{MD followed by Metropolis test}:%
  \footnote{ 
    $\delta_{T\calR}(z,\pi)$ is the delta function 
    associated with the symplectic volume form $d\Omega_\calR$. 
    $f_{\calR}$ denotes the MD integrator, 
    assumed to be both reversible and volume-preserving.
  } 
  \begin{align}
    P_{(2)}(z',\pi' \mid z,\pi) 
    &= \min\bigl(1,e^{-H(z',\pi') + H(z,\pi)}\bigr)\,
    \delta_{T\calR}((z',\pi') - f_\calR(z,\pi))
  \nonumber
  \\
    &\hspace{30mm}\mbox{for}~~
    (z',\pi') \neq (z,\pi).
  \label{wv_md}
  \end{align}%
\end{itemize}
The stochastic processes $P_{(1)}$ and $P_{(2)}$ 
satisfy the detailed balance condition \eqref{db_wv}, 
as shown in Ref.~\cite{Fukuma:2025gya} for more general cases. 

\begin{figure}[t]
  \centering
  \includegraphics[width=70mm]{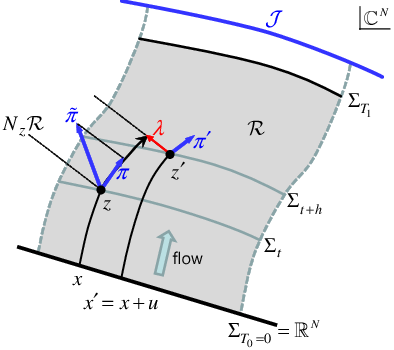}
  \caption{%
    RATTLE in WV-HMC 
    (figure adapted from Ref.~\cite{Fukuma:2023eru}).
  }
\label{fig:rattle_wv_wo_tpi_tR}
\end{figure}

MD on $T\calR$ of step size $\Delta s$ is implemented 
by repeating the RATTLE update of the following form 
(see Fig.~\ref{fig:rattle_wv_wo_tpi_tR}) 
\cite{Fukuma:2020fez,Fukuma:2023eru}:
\begin{align}
  \pi_{1/2} &= \pi - \Delta s\,\overline{\partial V(z)} - \Delta s\,\lambda,
\label{wv_rattle1}
\\
  z' &= z + \Delta s\,\pi_{1/2},
\label{wv_rattle2}
\\
  \pi' &= \pi_{1/2} - \Delta s\,\overline{\partial V(z')} - \Delta s\,\lambda'.
\label{wv_rattle3}
\end{align}%
Here, the Lagrange multipliers $\lambda$ and $\lambda'$ 
are chosen
so that $z' \in \calR$ and $\pi' \in T_{z'}\calR$, respectively. 
The force $-\overline{\partial V(z)}$ can be taken 
in the following form \cite{Fukuma:2020fez,Fukuma:2023eru}:
\begin{align}
  -\overline{\partial V(z)} 
  = -\frac{1}{2}\,\Bigl[ \xi + \frac{W'(t)}{\vev{\xi_n,\xi_n}}\,\xi_n \Bigr].
\end{align}%
The RATTLE update \eqref{wv_rattle1}--\eqref{wv_rattle3} is 
(i) exactly reversible, 
(ii) symplectic (and thus volume-preserving), 
and 
(iii) energy-conserving up to second order in $\Delta s$, 
$H(z',\pi') - H(z,\pi) = O(\Delta s^3)$.

The extent of the worldvolume $\calR$ in the flow-time direction 
can be effectively restricted to a finite interval $[T_0,T_1]$ 
by appropriately tuning the functional form of $W(t)$, 
which we take as follows \cite{Fukuma:2023eru}:
\begin{align}
  W(t) = 
  \begin{cases}  
    -\,\gamma(t-T_0) + c_0\,\bigl(e^{(t-T_0)^2/2d_0^2} - 1\bigr) 
    &  \mbox{\ for \ } t < T_0              
    \\
    -\,\gamma(t-T_0)                                             
    &  \mbox{\ for \ } T_0 \leq t \leq T_1  
    \\
    -\,\gamma(t-T_0) + c_1\,\bigl(e^{(t-T_1)^2/2d_1^2} - 1\bigr) 
    &  \mbox{\ for \ } t > T_1 .            
    \\
  \end{cases}
\label{Wt}
\end{align}%
Here, $c_0, d_0$ and $c_1, d_1$ determine the heights and penetration depths 
of the potential walls placed at $t \sim T_0$ and $t \sim T_1$, respectively. 
The tilt parameter $\gamma$ introduces a constant driving force 
to prevent configurations from accumulating at small flow times. 
These parameters are tuned, if necessary, 
so that configurations are distributed approximately uniformly 
along the flow-time direction.
The lower cutoff $T_0$ is chosen to ensure ergodicity 
on surfaces $\Sigma_t$ at $t \sim T_0$,
while the upper cutoff $T_1$ is set sufficiently large 
to suppress the oscillatory behavior of the integrand at $t \sim T_1$.

\subsection{Embedding GT-HMC into WV-HMC}
\label{sec:gt_into_wv}

In this subsection, 
we show that GT-HMC can be embedded into WV-HMC. 
The key point of the argument is that 
the subbundle $T\Sigma_t$ can be embedded into $T\calR$ 
in a manner that preserves the symplectic structure. 
In what follows, 
we denote $\Sigma_t$ simply by $\Sigma$ to simplify expressions. 

To this end, we parametrize the momentum $\hat{p} = (\hat{p}_\mu)$ 
as 
\begin{align}
  \hat{p}_0 \equiv e + \beta^c p_c,
  \quad
  \hat{p}_a \equiv p_a~~(a=1,\ldots,N).
\end{align}%
Noting that the metric $(\hat\gamma_{\mu\nu})$ 
and its inverse $(\hat\gamma^{\mu\nu})$ are given by 
\begin{align}
  (\hat\gamma_{\mu\nu})
  =
  \begin{pmatrix}
    \alpha^2 + \beta^c \beta_c & \beta_b \\
    \beta_a & \gamma_{ab}
  \end{pmatrix},
  \quad
  (\hat\gamma^{\mu\nu})
  =
  \begin{pmatrix}
    1/\alpha^2 & -\beta^b/\alpha^2 \\
    -\beta^a/\alpha^2 & \gamma^{ab} + \beta^a \beta^b /\alpha^2
  \end{pmatrix},
\end{align}%
we find that $\hat{p}^\mu = \hat\gamma^{\mu\nu} \hat{p}_\nu$ are given by 
\begin{align}
  \hat{p}^0 = \frac{e}{\alpha^2},
  \quad
  \hat{p}^a = p^a - \frac{e}{\alpha^2}\,\beta^a,
\end{align}%
which leads to the orthogonal decomposition 
of $\pi = \hat{E}_\mu\,\hat{p}^\mu \in T_z \calR$ 
into $\pi_n \in N_z \Sigma \cap T_z \calR$ and $\pi_v \in T_z \Sigma$ 
as 
\begin{align}
  \pi = \pi_n + \pi_v
  = \frac{e}{\alpha^2}\,\xi_n + E_a\, p^a, 
\label{pi_decomposition}
\end{align}%
where we have used $\xi - \beta^a E_a = \xi - \xi_v = \xi_n$. 
The decomposition \eqref{pi_decomposition} implies that 
MD restricted to the subbundle $T\Sigma \subset T\calR$ 
is characterized by the two constraints
\begin{align}
  z\in \Sigma,
  \quad
  e = 0,
\end{align}
because
\begin{align}
  (z,\pi) \in T\Sigma
  \quad\Leftrightarrow\quad
  z\in\Sigma,~~ \pi_n = 0
  \quad\Leftrightarrow\quad
  z\in \Sigma,
  \quad
  e = 0.
\end{align}%
Using the orthogonal decompositions \eqref{dz_decomposition} 
and \eqref{pi_decomposition},
the symplectic potential $a_\calR$ on $T\calR$ becomes 
\begin{align}
  a_\calR = \vev{\pi, dz}
  = (e + p_a \beta^a) dt + a_\Sigma,
\end{align}%
where $a_\Sigma = p_a dx^a$ is the symplectic potential on $T\Sigma$. 
We thus have the symplectic form $\omega_\calR = d a_\calR$ on $T\calR$ as 
\begin{align}
  \omega_\calR = [de + d(p_a \beta^a(t,x))] \wedge dt + \omega_\Sigma,
\end{align}%
where $\omega_\Sigma$ is the symplectic form on $T\Sigma$, 
$\omega_\Sigma = da_\Sigma = dp_a \wedge dx^a = d \vev{\pi,dz}$ 
[$(z,\pi) \in T\Sigma$]. 
This gives the symplectic volume form as 
\begin{align}
  d\Omega_\calR = \frac{\omega_\calR^{N+1}}{(N+1)!}
  = (de \wedge dt)\,\wedge\,d\Omega_\Sigma,
\end{align}%
where $d\Omega_\Sigma$ is the symplectic volume form on $T\Sigma$, 
$d\Omega_\Sigma = \omega_\Sigma^N/N! = \prod_a dx^a\,\prod_a dp_a$. 
This implies that 
the symplectic delta function on $T\calR$ can be written as 
\begin{align}
  \delta_{T\calR}(z,\pi) = \delta(t)\,\delta(e)\,\delta_{T\Sigma}(z,\pi),
\label{delta_wv}
\end{align}%
where $\delta_{T\Sigma}(z,\pi)$ is the symplectic delta function 
on $T\Sigma$. 

Now we are in a position to show that 
stochastic processes of GT-HMC can also be treated as those of WV-HMC. 
To this end, 
we first rewrite the subprocesses \eqref{gt_hb} and \eqref{gt_md} 
as transitions on $T\calR$: 
\begin{itemize}
\item
  \underline{Heat bath for $\pi$}:
  \begin{align} \vspace{-2ex}
    \tilde{P}_{(1)}(z',\pi' \mid z,\pi) 
    = e^{-(1/2) \vev{\pi',\pi'}}\,\delta(t'-t)\,\delta(e')\,\delta_\Sigma(z'-z),
  \label{wv_gt_hb}
  \end{align}%
  
\item
  \underline{MD followed by Metropolis test}:
  \begin{align}
    \tilde{P}_{(2)}(z',\pi' \mid z,\pi) 
    &= \min\bigl(1,e^{-H^\mathrm{GT}(z',\pi') + H^\mathrm{GT}(z,\pi)}\bigr)\,
    \delta(t'-t)\,\delta(e')\,
    \delta_{T\Sigma}((z',\pi') - f_\Sigma(z,\pi))
  \nonumber
  \\
    &= \min\bigl(1,e^{-H(z',\pi') + H(z,\pi)}\bigr)\,
    \delta_{T\calR}((z',\pi') - f_\Sigma(z,\pi))
  \nonumber
  \\
    &~~~~\mbox{for}~~
    (z',\pi') \neq (z,\pi).
  \label{wv_gt_md}
  \end{align}%
\end{itemize}
The subprocesses \eqref{wv_gt_hb} and \eqref{wv_gt_md} 
satisfy the detailed balance condition \eqref{db_wv} for WV-HMC. 
This is manifest for the first process \eqref{wv_gt_hb}. 
To prove this claim for the second process \eqref{wv_gt_md}, 
we need to make a careful argument 
with respect to the volume preservation. 
The point is that 
the delta function $\delta(t'-t)\,\delta(e')\,
\delta_{T\Sigma}((z',\pi') - f_\Sigma(z,\pi))$ 
is the symplectic delta function on $T\calR$ 
[see Eq.~\eqref{delta_wv}], 
and thus the time-reversed process 
$(z',-\pi') \to (z,-\pi)$ with the constraints $t'-t=0$ and $e'=0$ 
gives the unit Jacobian, 
and thus the standard argument of MD can be applied. 

We are now able to treat 
$\tilde{P}_{(1)}$ and $\tilde{P}_{(2)}$ 
as subprocesses for defining a stochastic process on $T\calR$. 
A simple implementation is to define 
$\tilde{P} \equiv \tilde{P}_{(2)} \tilde{P}_{(1)}$, 
and to consider a composition of $P$ (pure WV-HMC) 
and $\tilde{P}$ (embedded GT-HMC). 
This combined WV-HMC update 
yields configurations distributed according to $\propto e^{-H(z,\pi)}$, 
as long as the Markov chain includes $P$ sufficiently many times.

The embedding of GT-HMC into WV-HMC becomes particularly useful 
when the extent of $\calR$ in the flow-time direction is significantly small. 
In fact, as will be demonstrated in our study of the Hubbard model 
in subsequent sections, 
one can reduce the upper cutoff $T_1$ of the flow time 
by tuning a redundant, nonphysical parameter $\alpha$ \cite{Beyl:2017kwp} 
so as to suppress the sign problem on the original integration surface 
$\Sigma_0 = \bbR^N$. 
A smaller value of $T_1$ is generally preferable for simulations, 
because the flow equations need to be integrated only up to small flow times. 
However, this can introduce ergodicity issues for WV-HMC. 
Since the momentum $\pi$ is still generated isotropically in $\calR$ 
even when the worldvolume $\calR$ is a thin layer, 
it becomes difficult for configurations to efficiently explore the space. 
Moreover, irrespective of the thickness of the worldvolume, 
the step size $\Delta s$ of WV-HMC must be kept small 
to prevent configurations near the zeros of the Boltzmann weight 
from crossing into regions unreachable from $\Sigma_0$. 

In contrast, GT-HMC is uniquely suited to overcome these specific inefficiencies. 
Although GT-HMC suffers from its own ergodicity problem 
due to the zeros of the Boltzmann weight, 
it enables more efficient exploration within the allowed regions. 
Additionally, $\Delta s$ of GT-HMC can be set to a larger value than in WV-HMC 
because configurations in GT-HMC feel repulsion from the zeros 
more effectively (only along the direction tangent to the deformed surface). 
The combined algorithm thus provides a way 
to enhance the ergodicity of WV-HMC 
even when the worldvolume is a thin layer. 
Note that the MD step size $\Delta s$ 
can be set differently for WV-HMC and GT-HMC.%
\footnote{ 
  It can even be varied for GT-HMC subprocesses 
  depending on the flow time, 
  although we do not pursue this possibility in this paper. 
} 

\section{Application to the Hubbard model}
\label{sec:application}

In this section, 
we prepare the necessary ingredients 
for applying the pure and combined WV-HMC algorithms 
to the doped Hubbard model.

\subsection{The Hubbard model}
\label{sec:model}

The Hubbard model on a $d$-dimensional spatial lattice 
is defined by the following Hamiltonian 
(including the chemical potential term): 
\begin{align}
  \hat H^\mathrm{org}_\mu
  &= \hat H - \mu \hat N
\nonumber
\\
  &\equiv
  -\sum_{\bx, \by} t_{\bx \by} \sum_{\sigma=\uparrow,\downarrow} 
    c_{\bx, \sigma}^\dag c_{\by, \sigma}
  + U      \sum_{\bx} n_{\bx, \uparrow} n_{\bx, \downarrow} 
  - \mu    \sum_{\bx} (n_{\bx, \uparrow} + n_{\bx, \downarrow}) .
\label{hamiltonian_org}
\end{align}%
Here,  $c_{\bx, \sigma}$ and $c_{\bx, \sigma}^\dag$ denote  
the annihilation and creation operators, respectively, 
of an electron with spin $\sigma\,(=\uparrow,\,\downarrow)$ 
at site $\bx=(x_i)$ $(i=1,\ldots,d)$, 
and $n_{\bx, \sigma} \equiv c_{\bx, \sigma}^\dag c_{\bx, \sigma}$.  
$t = (t_{\bx \by})$ is the hopping matrix, 
where $t_{\bx\by} = t\,(>0)$ if $\bx$ and $\by$ are nearest neighbors,
and $t_{\bx\by} = 0$ otherwise.
$U$ is the on-site repulsion strength, 
and $\mu$ is the chemical potential associated with the number operator 
$\hat N = \sum_\bx \sum_\sigma n_{\bx,\sigma}$.
We assume that the model is defined on a periodic, bipartite square lattice 
of linear size $L_s$, 
so that the spatial volume is given by $V_d \equiv L_s^d$. 

To make the real-valuedness of the bosonized action 
(introduced below) manifest at half filling, 
we perform a particle-hole transformation on the down-spin component 
and write 
\begin{align}
  a_\bx \equiv c_{\bx\uparrow},\quad
  b_\bx \equiv (-1)^\bx c^\dag_{\bx\downarrow},
\end{align}%
where $(-1)^\bx\equiv (-1)^{\sum_i x_i}$ denotes the parity of site $\bx$.
Under this transformation, 
the Hamiltonian \eqref{hamiltonian_org} becomes 
(up to an additive constant $-\mu V_d$):
\begin{align}
  \hat H_\mu 
  \equiv H^\mathrm{org}_\mu - \mu V_d
  = -\sum_{\bx, \by} 
  t_{\bx \by}\, ( a_\bx^\dag a_\by + b_\bx^\dag b_\by )
  + \frac{U}{2}\, \sum_{\bx} ( n^a_\bx - n^b_\bx )^2
  - \tilde \mu\, \sum_{\bx} ( n^a_\bx - n^b_\bx ) ,
\end{align}%
where $n^a_\bx\ \equiv a_\bx^\dag a_\bx$ 
and $n^b_\bx\ \equiv b_\bx^\dag b_\bx$, 
and the shifted chemical potential is defined as  
\begin{align}
  \tilde \mu \equiv \mu - \frac{U}{2}.
\end{align}%
The point $\mu=U/2$ (i.e., $\tilde\mu=0$) corresponds to half filling, 
where $\vev{ n_{\bx,\uparrow} + n_{\bx,\downarrow} } = 1$ 
(i.e., $\vev{ n^a_\bx - n^b_\bx } = 0$). 

Following Ref.~\cite{Beyl:2017kwp}, 
we introduce the redundant parameter $\alpha$ $(0\leq\alpha\leq 1)$ as%
\footnote{ 
  The equality directly follows from the identity 
  $ (n^a_\bx + n^b_\bx - 1)^2 = -(n^a_\bx - n^b_\bx)^2 + 1$ 
  (note that $(n^{a/b}_\bx)^2 = n^{a/b}_\bx$)
  \cite{Beyl:2017kwp}. 
} 
\begin{align}
  (n^a_\bx - n^b_\bx)^2 = \alpha (n^a_\bx - n^b_\bx)^2
  - (1-\alpha) (n^a_\bx + n^b_\bx -1)^2 + 1 - \alpha.
\label{alpha}
\end{align}%
We then complete the square 
by using two auxiliary variables (Hubbard-Stratonovich variables): 
\begin{align}
  &e^{-(\alpha\epsilon U/2)\,(n^a - n^b)^2 
      + ((1-\alpha)\epsilon U/2)\,(n^a + n^b - 1)^2 
      - (1-\alpha)\epsilon U/2}
\nonumber
\\
  &= \int dA dB\, e^{-(1/2)(A^2 + B^2)}\,
  e^{ [ i c_0 A + c_1 B - c_1^2] \, n^a} \,
  e^{ [-i c_0 A + c_1 B - c_1^2] \, n^b} 
\end{align}%
with 
\begin{align}
  c_0 \equiv \sqrt{\alpha\epsilon U}, \quad
  c_1 \equiv \sqrt{(1-\alpha)\epsilon U} . 
\end{align}%
We decompose the inverse temperature $\beta$ into $N_t$ time slices 
and introduce a spacetime lattice of volume 
$V_{d+1} \equiv N_t V_d = N_t\times L_s^d$, 
whose coordinates are labeled by $x = (\ell,\bx)$ 
$(\ell = 1,\ldots,N_t)$.   
Then, the grand canonical partition function is given as follows 
(see Ref.~\cite{Fukuma:2025uzg} for the derivation):  
\begin{align}
  Z = \int dA\,dB\,e^{-S(A,B)}
  = \int dA\,dB\,e^{-(1/2)\sum_x (A_x^2 + B_x^2)}\,\det D_a(A,B)\,\det D_b(A,B).
\label{path_int_AB}
\end{align}%
Here, $A = (A_x)$ and $B = (B_x)$ are scalar fields on the spacetime lattice, 
and we have introduced 
$V_{d+1} \times V_{d+1}$ matrices 
$t = (t_{xy})$ and $\Lambda_0 = ((\Lambda_0)_{xy})$ 
with indices $x = (\ell, \bx)$ and $y= (m, \by)$ (we reuse the symbol $t$) 
as 
\begin{align}
  t_{xy} &\equiv \delta_{\ell m}\,t_{\bx\by},
\\
  (\Lambda_0)_{xy} &\equiv 
   \left\{ \begin{array}{ll}
      \delta_{\ell+1,m}\,\delta_{\bx\by} & (\ell < N_t) \\
      - \delta_{1,m}\,\delta_{\bx\by}    & (\ell = N_t).
   \end{array}\right. 
\end{align}%
$h_{a/b} = ((h_{a/b})_x)$ are diagonal matrices with 
\begin{align}
  (h_{a/b})_x = e^{\pm (\epsilon\tilde\mu + i c_0 A_x) + c_1 B_x - c_1^2}, 
\end{align}%
and $D_{a/b}$ are fermion matrices, 
\begin{align}
  D_{a/b}(A,B) \equiv 
  h_{a/b} - e^{-\epsilon t}\,\Lambda_0.
\end{align}%
We employ the symmetric Trotter decomposition, 
which agrees with the continuum transfer matrix $e^{-\epsilon \hat{H}_\mu}$ 
to second order in $\epsilon$, 
and accordingly, 
we expand $D_{a/b}$ to the same order:%
\footnote{ 
  Note that 
  $\Lambda_0 = 1 + O(\epsilon)$ holds only for thermalized configurations 
  and should not be used as a general estimate. 
} 
\begin{align}
  D_{a/b} = h_{a/b} - \Lambda_0 + \epsilon t \Lambda_0
    - \frac{\epsilon^2}{2}\,t^2 \Lambda_0.
\end{align}%

As discussed in Ref.~\cite{Fukuma:2025uzg}, 
we have the identity $D_b = \overline{D_a}$ at half filling ($\tilde\mu = 0$), 
which leads to $\det D_a\, \det D_b = |\det D_a|^2$.
This implies that the path integral is free from the sign problem at half filling.
Moreover, we expect the sign problem to remain mild 
when $D_b \approx \overline{D_a}$, 
which is realized 
when $\alpha$ takes a small value. 
However, choosing $\alpha$ too small introduces ergodicity issues 
due to the appearance of zeros of $\det D_{a/b}$ 
on or near the original configuration space $\Sigma_0$, 
as investigated in detail in Ref.~\cite{Beyl:2017kwp}.
Thus, there is an optimal value for $\alpha$ 
that reduces (but does not completely remove) the sign problem on $\Sigma_0$ 
without introducing ergodicity problems.

\subsection{Observables}
\label{sec:wv1_observables}

When the Trotter number $N_t$ is held fixed, 
the parameters $\beta$ and $\beta\mu$ enter the action 
only through $\epsilon$ 
and $\epsilon\tilde\mu = \epsilon\mu - \epsilon U/2$, respectively. 
We define 
the number density operator $n$ 
and the energy density operator $e$ 
as follows \cite{Fukuma:2025uzg}:
\begin{align}
  n(A,B) 
  &\equiv 
  - \frac{1}{V_{d+1}}\,
    \frac{\partial S(A,B)}{\partial (\epsilon\mu)}\Bigr|_{\epsilon}
  + 1
  = - \frac{1}{V_{d+1}}\,
    \frac{\partial S(A,B)}{\partial (\epsilon\tilde\mu)}\Bigr|_{\epsilon}
  + 1 ,
\label{number_density}
\\
  e(A,B) 
  &\equiv 
  \frac{\partial S(A,B)}{\partial \epsilon}\Bigr|_{\epsilon\mu}
  = \frac{1}{V_{d+1}}\,\biggl[
  \frac{\partial S(A,B)}{\partial \epsilon}\Bigr|_{\epsilon\tilde\mu}
  - \frac{U}{2}\,
  \frac{\partial S(A,B)}{\partial (\epsilon\tilde\mu)}\Bigr|_{\epsilon}
  \biggr] .
\label{energy_density}
\end{align}%
Their expectation values can be estimated via the path integral 
and are expected to agree with the continuum expectation values 
of $\hat N/V_d$ and $\hat H/V_d$ 
up to $O(\epsilon^2)$ corrections: 
\begin{align}
  \langle n \rangle &\equiv 
  \frac{1}{V_{d+1}}\,\frac{\int (dA\,dB)\,e^{-S(A,B)}\,n(A,B)}
       {\int (dA\,dB) e^{-S(A,B)}}
  = \frac{1}{V_d}\,\frac{\tr\, e^{-\beta (\hat H - \mu \hat N)}\,\hat N }
                        {\tr\, e^{-\beta (\hat H - \mu \hat N)} }
    + O(\epsilon^2) ,
\label{scaling_n}
\\
 \langle e \rangle 
 &\equiv 
 \frac{1}{V_{d+1}}\,\frac{\int (dA\,dB)\,e^{-S(A,B)}\,e(A,B)}
 {\int (dA\,dB) e^{-S(A,B)}}
 = \frac{1}{V_d}\,\frac{\tr\, e^{-\beta (\hat H - \mu \hat N)}\,\hat H }
 {\tr\, e^{-\beta (\hat H - \mu \hat N)} }
 +  O(\epsilon^2). 
\label{scaling_e}
\end{align}%

\section{Numerical tests}
\label{sec:numerical_tests}

In this section, 
we apply the combined algorithm to the two-dimensional doped Hubbard model 
on a spatial lattice of size $L_s \times L_s = 8 \times 8$ 
with the following parameters:  
hopping amplitude $t = 1.0$, 
inverse temperature $\beta = 6.4$, 
repulsion strength $U = 8.0$, 
and a varying chemical potential $\mu$.%
\footnote{ 
  In the common notation in condensed matter physics, 
  these parameters correspond to 
  $T/t = 1/(t\beta) = 1/6.4 \simeq 0.156$ and $U/t = 8.0$. 
} 

We first show that, 
in parameter regions suffering from a severe sign problem, 
the pure and combined algorithms 
at Trotter step $\epsilon = 0.32$ (corresponding to Trotter number $N_t = 20$)
yield the same results within statistical errors. 
We then compute the number and energy densities 
at $\epsilon = 0.27,\,0.29,\,0.32,\,0.36$ 
(corresponding to Trotter numbers $N_t = 24,\,22,\,20,\,18$), 
and extrapolate the obtained results 
to the zero Trotter step limit ($\epsilon \to 0$) 
using the scaling relations \eqref{scaling_n} and \eqref{scaling_e}.

\subsection{Tuning of $\alpha$}
\label{sec:tuning_alpha}

\begin{figure}[ht]
  \centering
  \includegraphics[width=70mm]{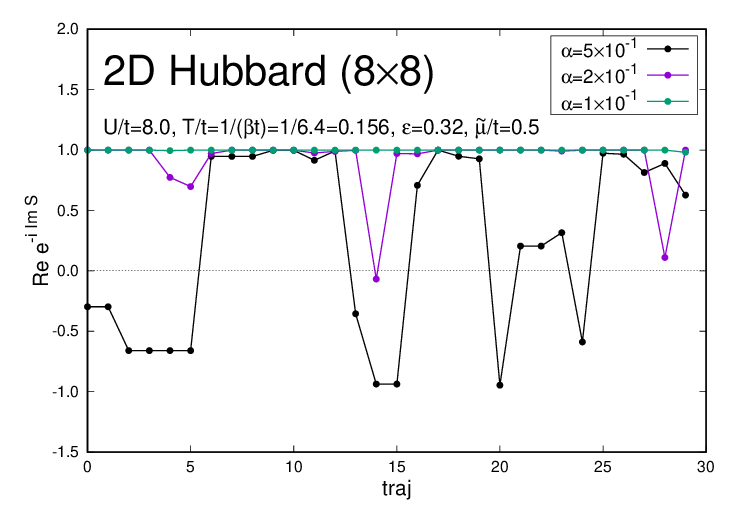}
  \includegraphics[width=70mm]{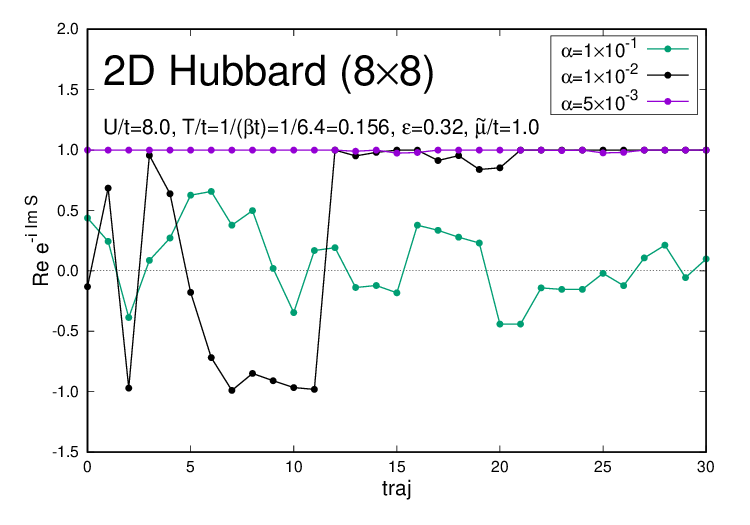}
  \includegraphics[width=70mm]{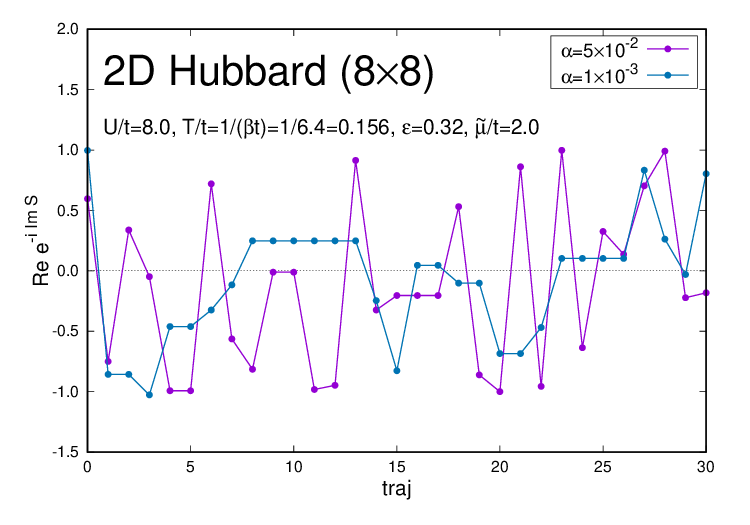}
  \includegraphics[width=70mm]{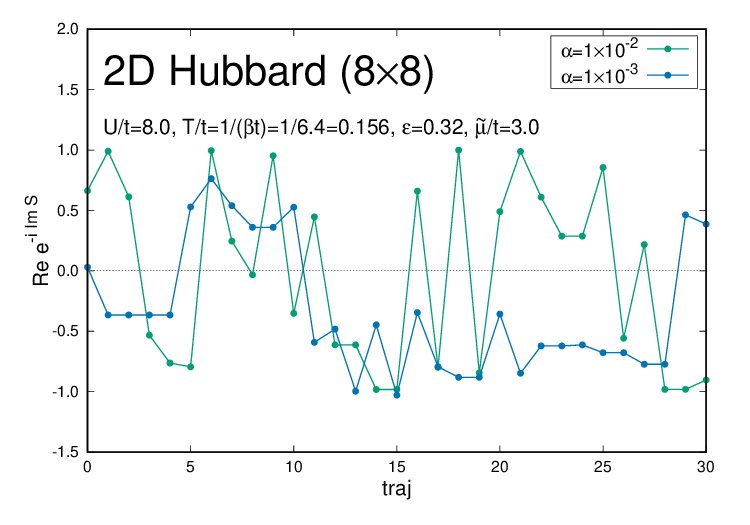}
  \includegraphics[width=70mm]{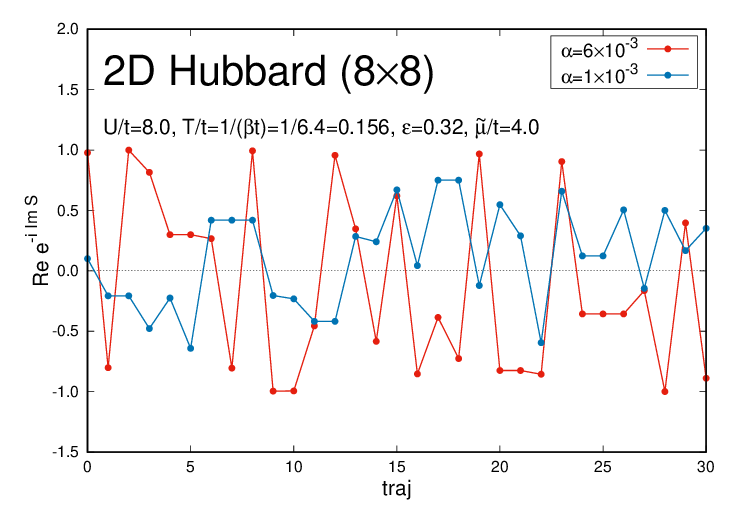}
  \includegraphics[width=70mm]{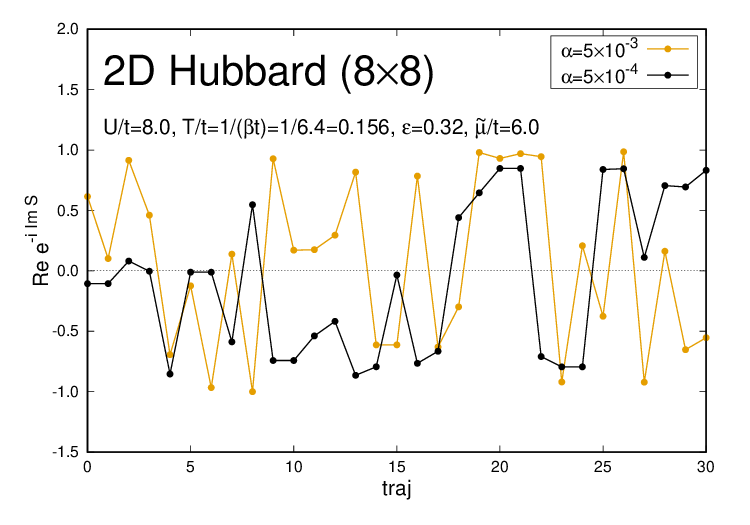}
  \caption{%
    Histories of the phase factor on $\Sigma_0$ 
    for various values of $\alpha$ 
    at $\tilde\mu = 0.5,\,1.0,\,2.0,\,3.0,\,4.0,\,6.0$. 
  }
\label{fig:alpha_dependence}
\end{figure}

Although the tuning of $\alpha$ was previously performed 
in Ref.~\cite{Fukuma:2025uzg},
we re-tune it here to address the cases 
where the maximum flow time $T_1$ is set as small as possible. 
This requires identifying the smallest value of $\alpha$ 
that still avoids ergodicity issues. 
Figure~\ref{fig:alpha_dependence} shows 
the histories of the phase factor on $\Sigma_0$ 
for shifted chemical potentials 
$\tilde\mu = 0.5,\,1.0,\,2.0,\,3.0,\,4.0,\,6.0$. 
To ensure the absence of ergodicity issues, 
we impose the following criterion: 
the length of every plateau must be shorter than 10 trajectories. 
The selected values of $\alpha$ based on this criterion 
are summarized in Table~\ref{table:alpha_tuned_combo}. 
These values are smaller than those used in pure WV-HMC \cite{Fukuma:2025uzg}
except for $\tilde\mu=0.5$ and $1.0$.
\begin{table}[ht]
  \centering
  \begin{tabular}{|c|c|c|c|c|} \hline
    $\tilde{\mu}$  &  0.5  &   1.0    &  1.5 -- 5.0            &   6.0 -- 9.0
    \\ \hline
    $\alpha$       &  0.5  &   0.1    &  $1.0 \times 10^{-3}$  &  $5.0 \times 10^{-4}$
    \\ \hline
  \end{tabular}
  \caption{%
    Tuned values of $\alpha$ used in the combined WV-HMC algorithm.
  }
\label{table:alpha_tuned_combo}
\end{table}%

\begin{figure}[ht]
  \centering
  \includegraphics[width=70mm]{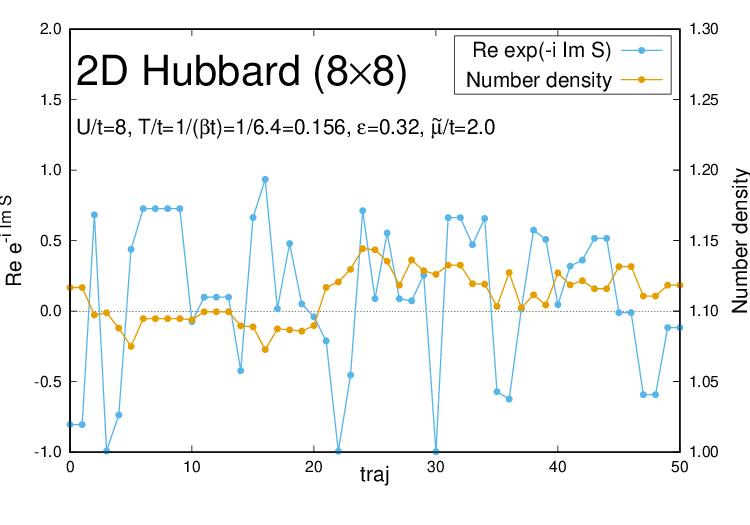}
  \hspace{3mm}
  \includegraphics[width=70mm]{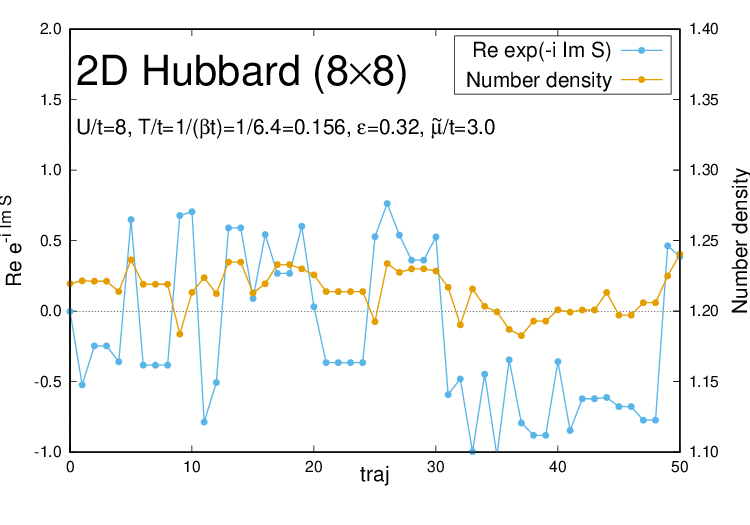}
  \\
  \includegraphics[width=70mm]{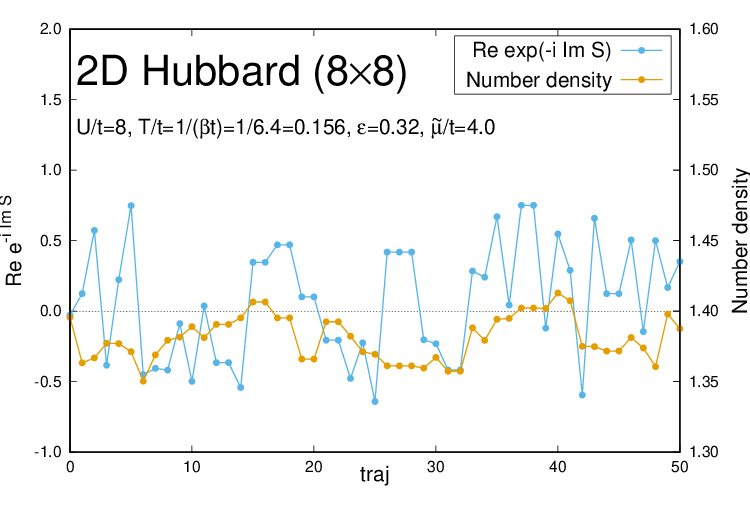}
  \hspace{3mm}
  \includegraphics[width=70mm]{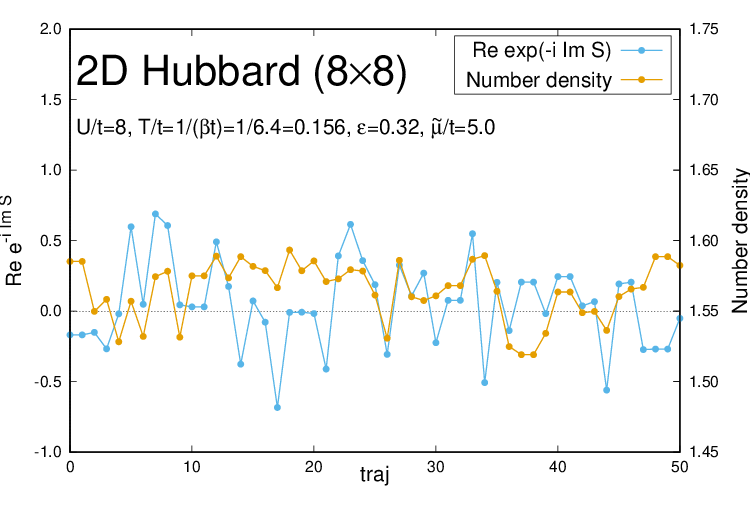}
  \\
  \includegraphics[width=70mm]{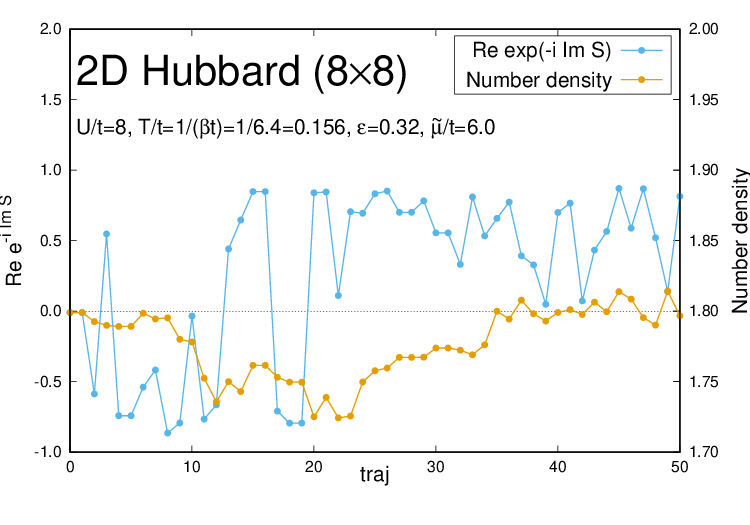}
  \hspace{3mm}
  \includegraphics[width=70mm]{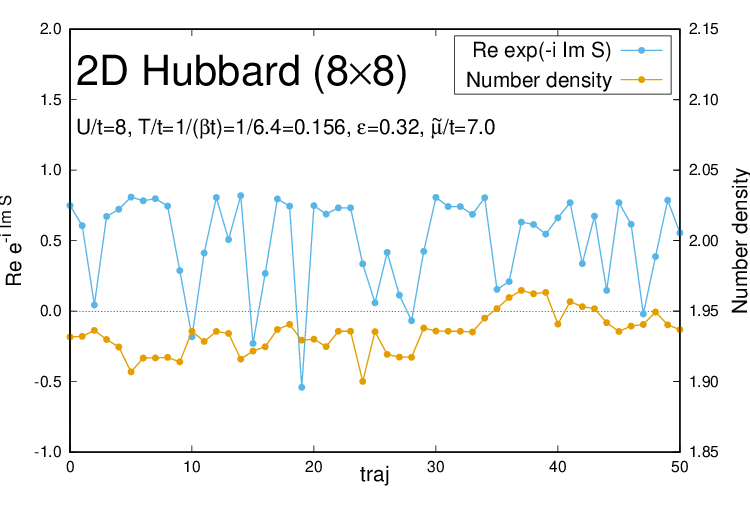}
  \\
  \includegraphics[width=70mm]{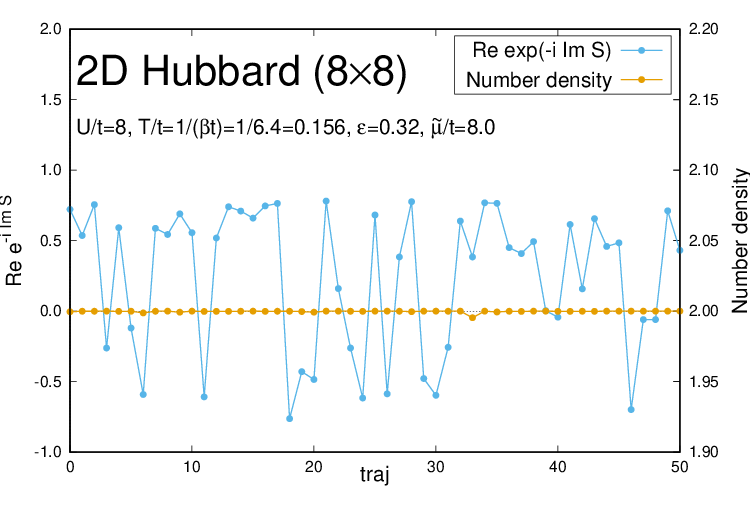}
  \hspace{3mm}
  \includegraphics[width=70mm]{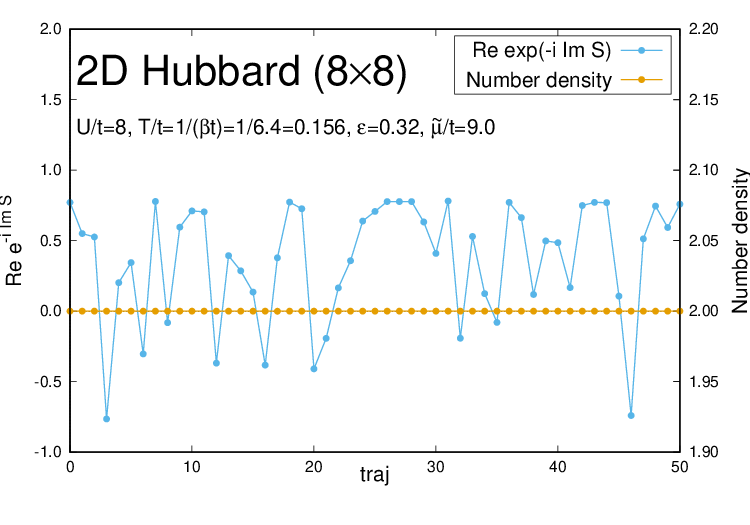}
  \caption{%
    Histories of the phase factor and the number density on $\Sigma_0$ 
    for various values of $\tilde\mu$ 
    obtained using the tuned values of $\alpha$ 
    shown in Table~\ref{table:alpha_tuned_combo}.
  }
\label{fig:alpha_tuned}
\end{figure}

Figure~\ref{fig:alpha_tuned} shows 
the histories of the phase factor and the number density on $\Sigma_0$ 
obtained using the tuned values of $\alpha$. 
The rapid fluctuations observed in both quantities, 
together with the absence of long plateaus, 
indicate that ergodicity issues are effectively suppressed.
%

\subsection{Tuning of $T_1$ and setup of other parameters}
\label{sec:tuning_T1}

\begin{figure}[t]
  \centering
  \includegraphics[width=75mm]{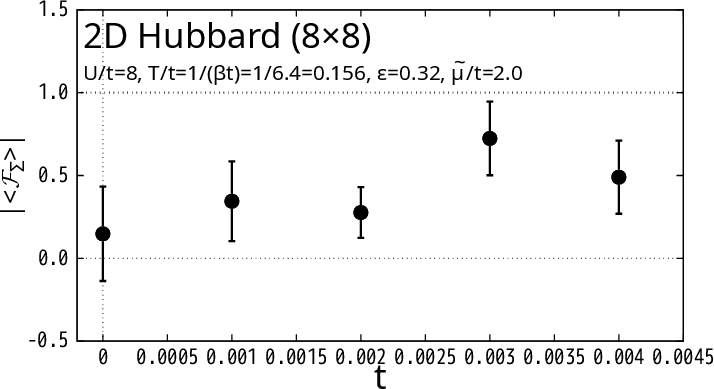}
  \hspace{3mm}
  \includegraphics[width=75mm]{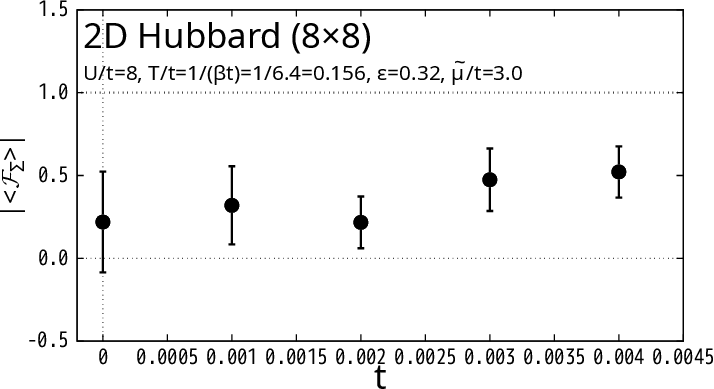}
  \\
  \includegraphics[width=75mm]{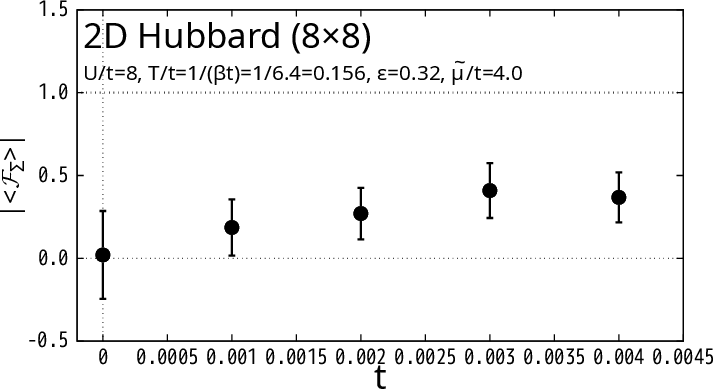}
  \hspace{3mm}
  \includegraphics[width=75mm]{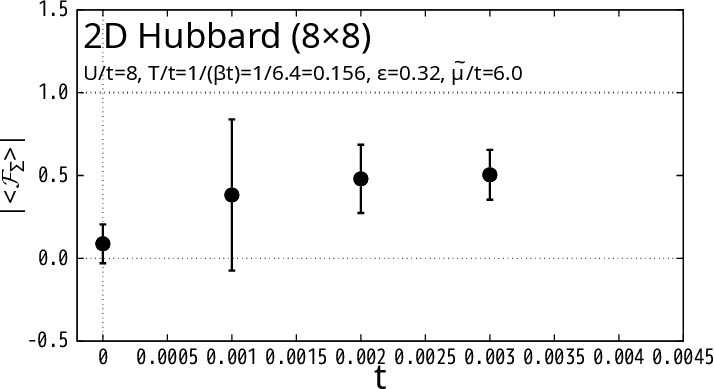}
  \caption{%
    Flow-time dependence of the average reweighting factor 
    for various values of $\tilde\mu$.
  }
\label{fig:hubbard_T1}
\end{figure}

Figure~\ref{fig:hubbard_T1} shows 
the average reweighting factor $\vev{\calF_\Sigma}$ [Eq.~\eqref{rewtfactor_GT}] 
on the deformed surface $\Sigma_t$ 
as a function of the flow time $t$, 
computed using GT-HMC. 
We observe that, for each value of $\tilde\mu$, 
the average becomes statistically distinguishable from zero 
at the two-sigma level 
for sufficiently large flow times. 
Based on these observations, 
we choose the upper cutoff $T_1$ 
as listed in Table~\ref{table:T1_combo}. 
\begin{table}[ht]
  \centering
  \begin{tabular}{|c|c|c|c|} \hline
    $\tilde{\mu}$ & 0.5 -- 2.2           & 2.5 -- 5.0           & 6.0 -- 9.0
    \\ \hline
    $T_1$         & $4.0 \times 10^{-3}$ & $8.0 \times 10^{-3}$ & $4.0 \times 10^{-3}$
    \\ \hline
  \end{tabular}
\caption{%
  Values of $T_1$ used in the combined WV-HMC algorithm.}
\label{table:T1_combo}
\end{table}%

The weight function parameters [see Eq.~\eqref{Wt}] 
are set as follows: 
$\gamma = 0$, $c_0 = c_1 = 0.01$, and $d_0 = d_1 = 2.0\times 10^{-3}$. 
The simulations are carried out using a combined update 
that consists of 
two sets of embedded GT-HMC 
(each trajectory consisting of 25 MD steps  
with $\Delta s = 4.0 \times 10^{-2}$) 
followed by 
one set of pure WV-HMC 
(each trajectory consisting of 25 MD steps 
with $\Delta s = 4.0 \times 10^{-4}$). 
Observables are measured after each combined update, 
and statistical errors are estimated 
using the blocked jackknife method with adjusted bin sizes.

\subsection{Comparison of the combined algorithm with pure WV-HMC}

\begin{figure}[ht]
  \centering
  \includegraphics[width=70mm]{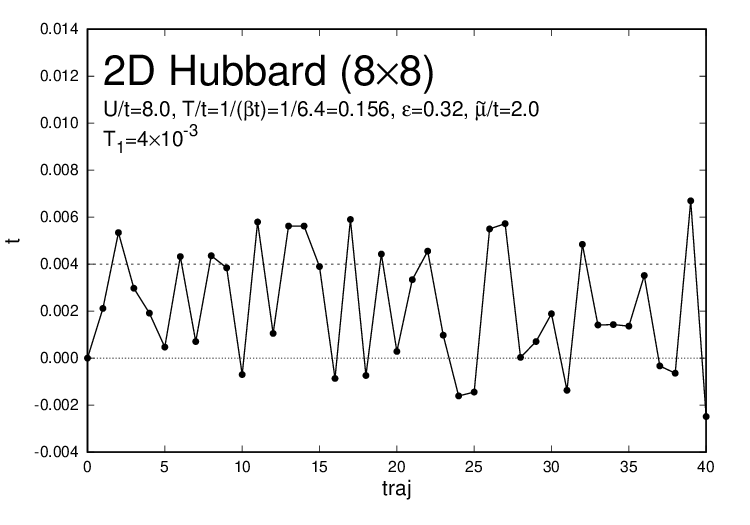}
  \includegraphics[width=70mm]{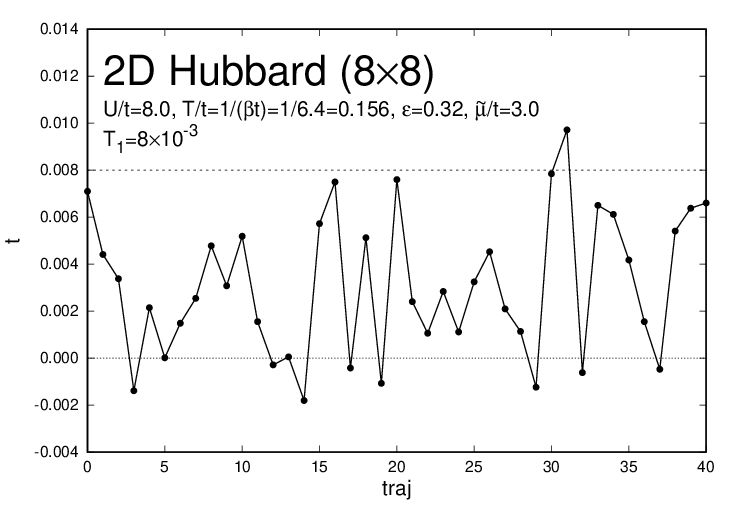}
  \includegraphics[width=70mm]{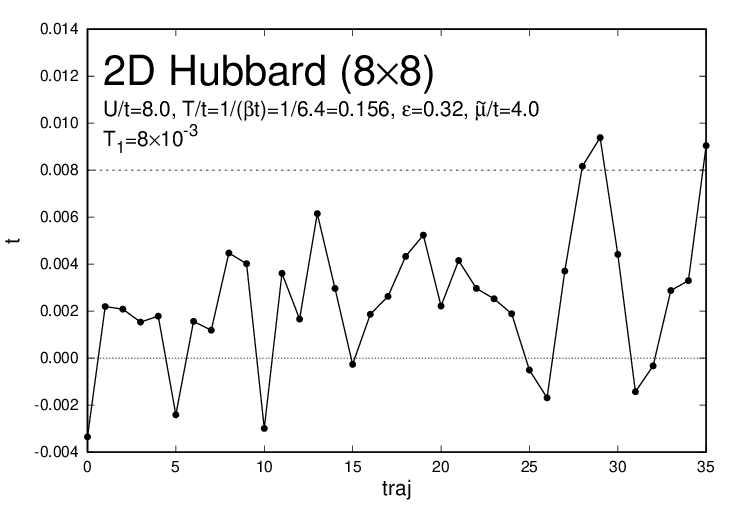}
  \includegraphics[width=70mm]{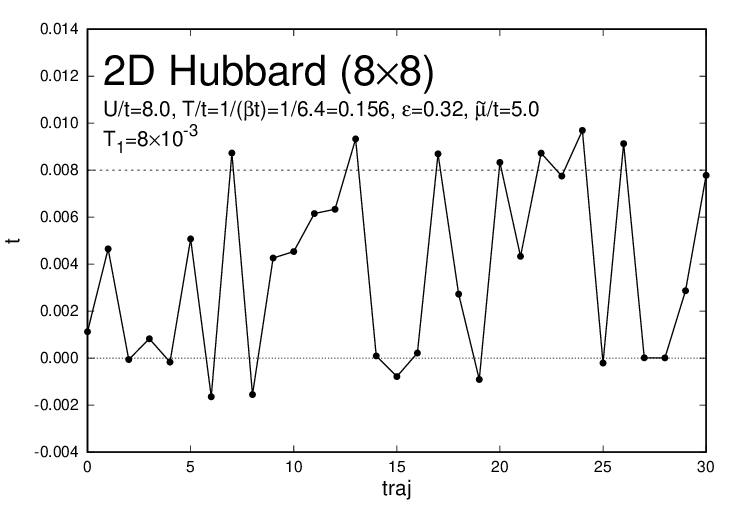}
  \includegraphics[width=70mm]{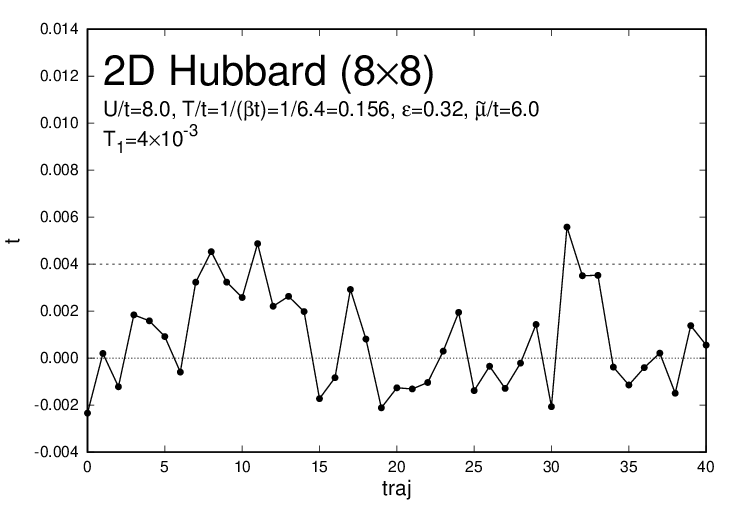}
  \includegraphics[width=70mm]{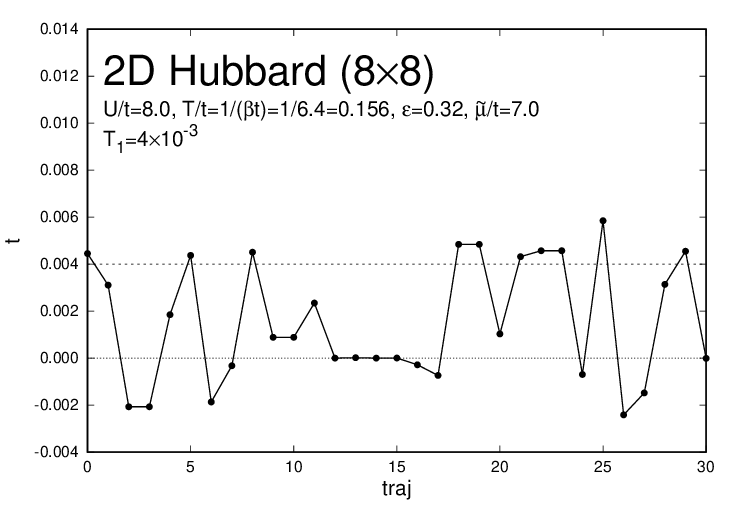}
  \includegraphics[width=70mm]{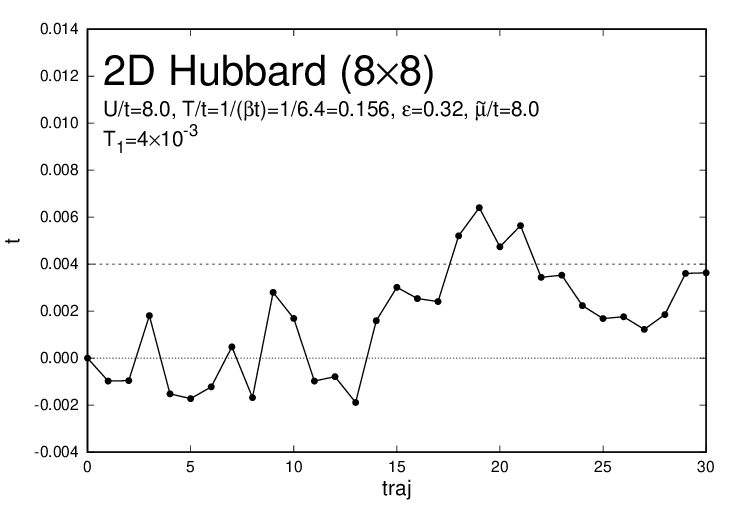}
  \includegraphics[width=70mm]{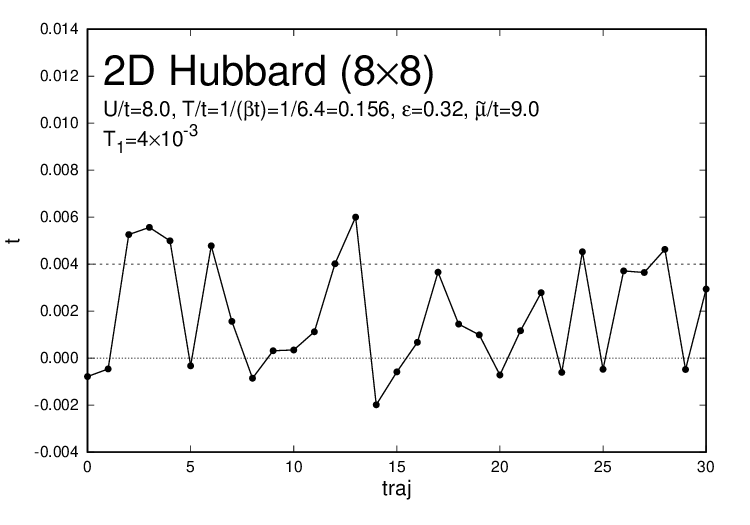}
  \caption{%
    Histories of the flow time in the combined WV-HMC algorithm.
  }
\label{fig:tflow_8x8a}
\end{figure}

\begin{figure}[ht]
  \centering
  \includegraphics[width=100mm]{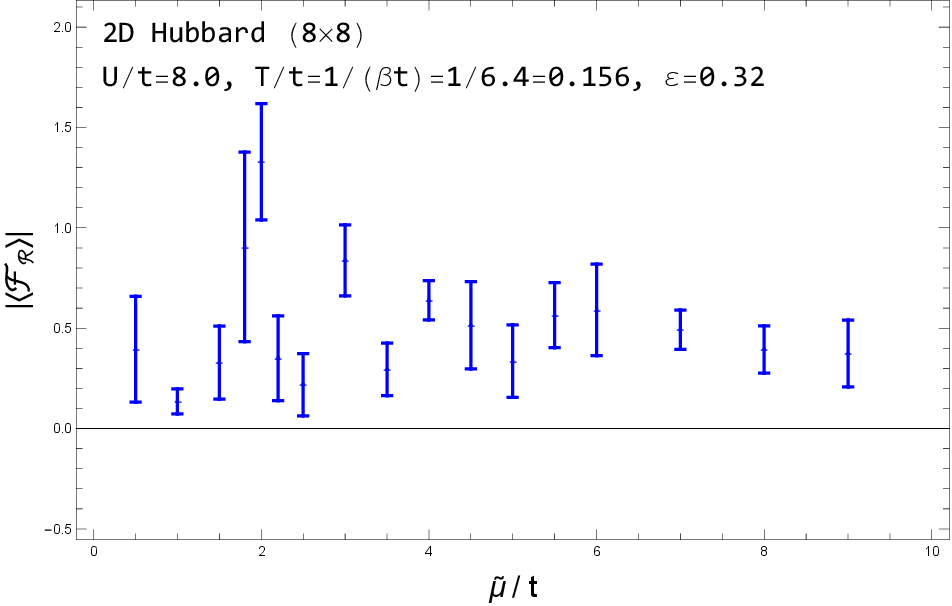}
  \caption{%
    Average reweighting factor obtained using the combined WV-HMC algorithm.
  }
\label{fig:average_reweighting_factor_8x8a}
\end{figure}

Figure~\ref{fig:tflow_8x8a} shows the histories of the flow time 
obtained using the combined WV-HMC algorithm. 
We observe that the flow time fluctuates rapidly 
within the short interval $[T_0,\,T_1]$, 
over which the corresponding worldvolume forms a thin layer. 
Figure~\ref{fig:average_reweighting_factor_8x8a} displays 
the average reweighting factor $\vev{\calF_\calR}$ [Eq.~\eqref{rewtfactor_R}] 
computed using the combined WV-HMC algorithm 
with the tuned values of $\alpha$ and $T_1$. 
These averages are statistically distinguishable from zero, 
indicating the reliability of the Monte Carlo sampling for these parameter choices. 

\begin{figure}[ht]
  \centering
  \includegraphics[width=140mm]
    {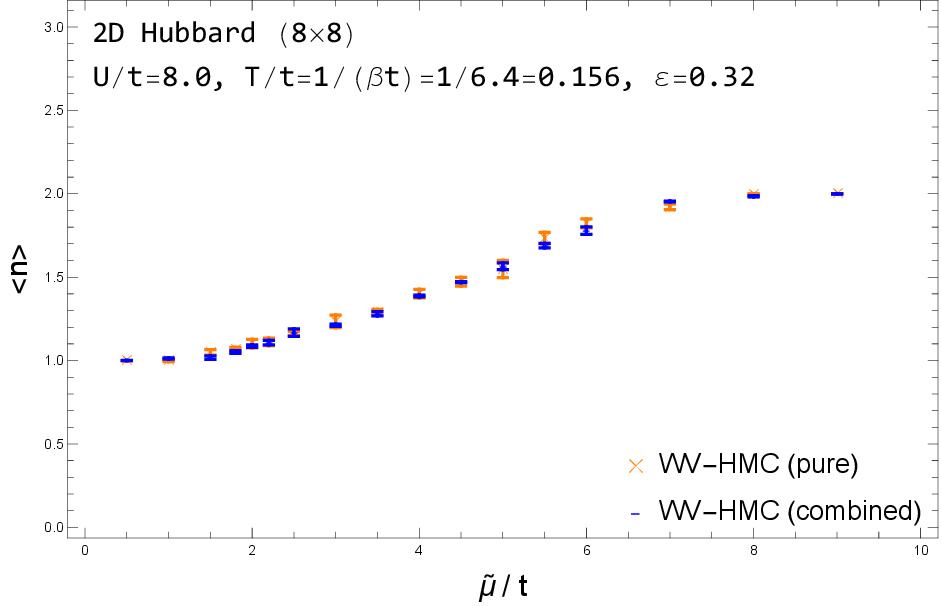}
  \vspace{3ex}\\
  \includegraphics[width=140mm]
    {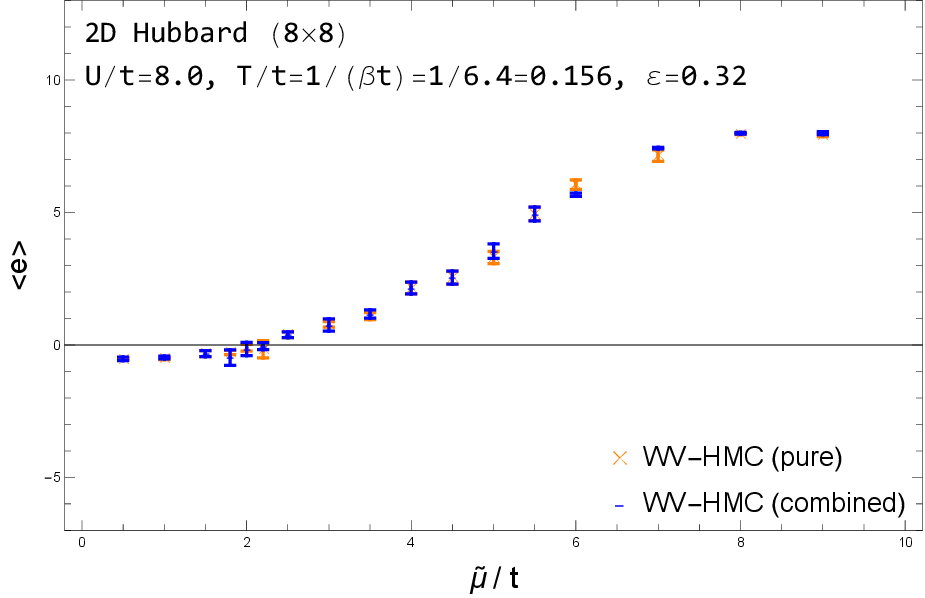}
  \caption{%
    Number density $\vev{n}$ and energy density $\vev{e}$ 
    at $\epsilon = 0.32$  
    obtained using the pure and combined WV-HMC algorithms. 
    The results of pure WV-HMC are taken from Ref.~\cite{Fukuma:2025uzg}, 
    where the upper cutoff was set to $T_1 = 0.1$.
  }
\label{fig:observables_20x8x8}
\end{figure}

Figure~\ref{fig:observables_20x8x8}  
shows the number and energy densities 
obtained using the pure and combined WV-HMC algorithms.%
\footnote{ 
  See Ref.~\cite{Fukuma:2025uzg} 
  for a comparison with results obtained using ALF \cite{Bercx:2017pit,ALF:2020tyi}. 
} 
The results of pure WV-HMC are taken from Ref.~\cite{Fukuma:2025uzg}, 
where the upper cutoff was set to $T_1 = 0.1$, 
which is sufficiently large to avoid ergodicity issues. 
The results from the pure and combined algorithms 
agree within statistical errors, 
demonstrating that 
the embedding indeed works for the tuned values of $\alpha$ and $T_1$. 
%

\subsection{Continuum limit in the temporal direction}

The combined WV-HMC algorithm allows simulations on larger spacetime lattices. 
In this subsection, 
we extrapolate to the zero Trotter step limit ($\epsilon \to 0$)  
at fixed inverse temperature $\beta = 6.4$ 
and hopping amplitude $t = 1.0$ 
(corresponding to the fixed ratio $T/t = 1/6.4 \simeq 0.156$). 

\begin{figure}[ht]
  \centering
  \includegraphics[width=52mm]{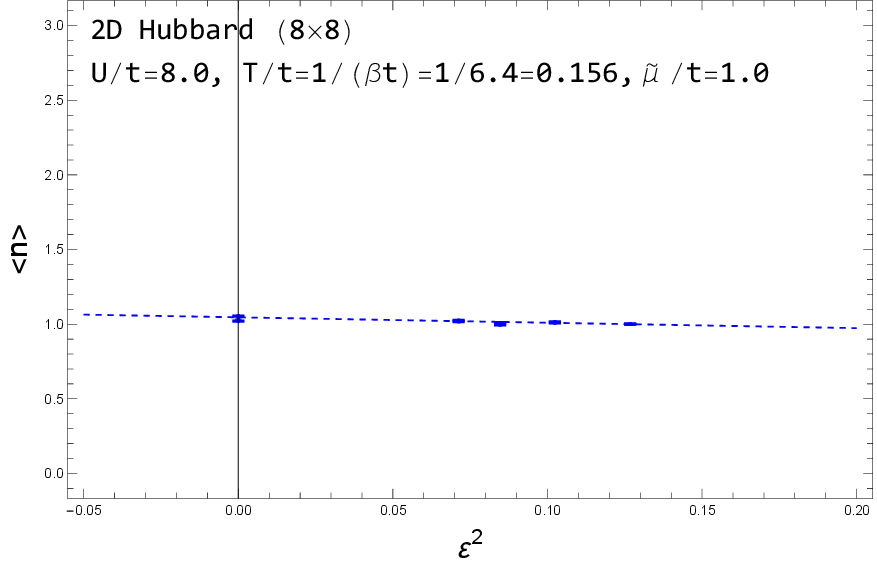}
  \hspace{1mm}
  \includegraphics[width=52mm]{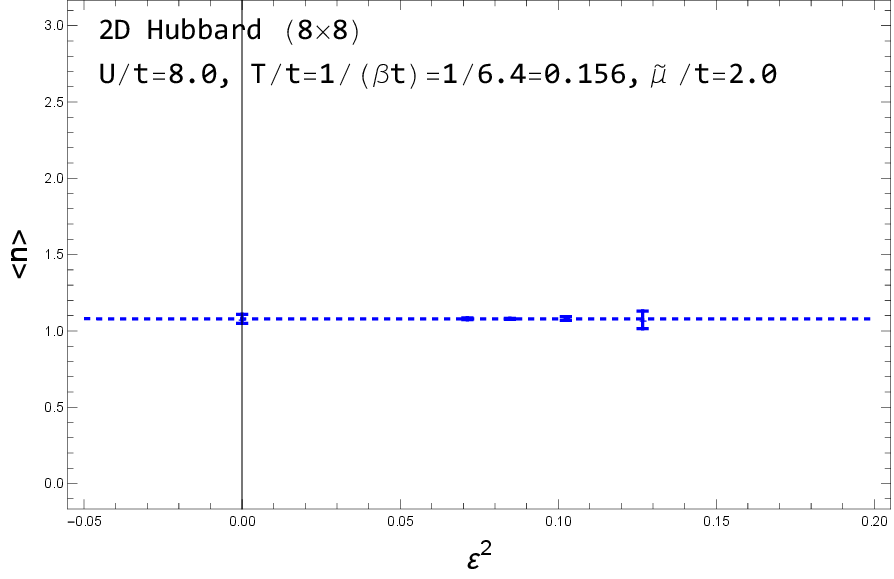}
  \hspace{1mm}
  \includegraphics[width=52mm]{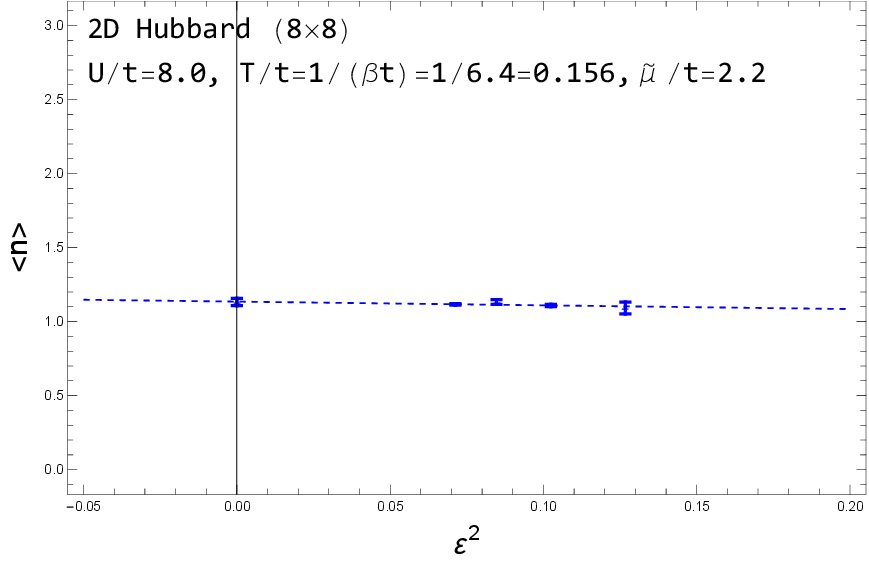}
  \vspace{3ex}
  \\
  \includegraphics[width=52mm]{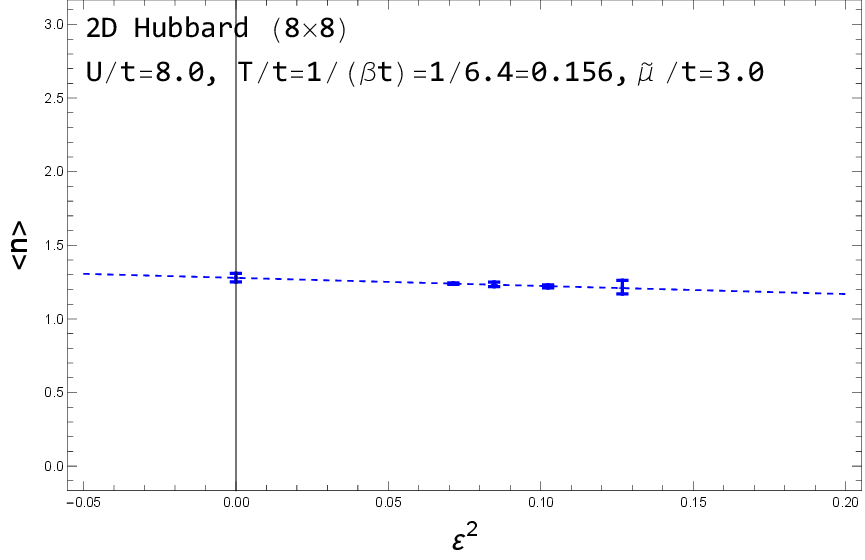}
  \hspace{1mm}
  \includegraphics[width=52mm]{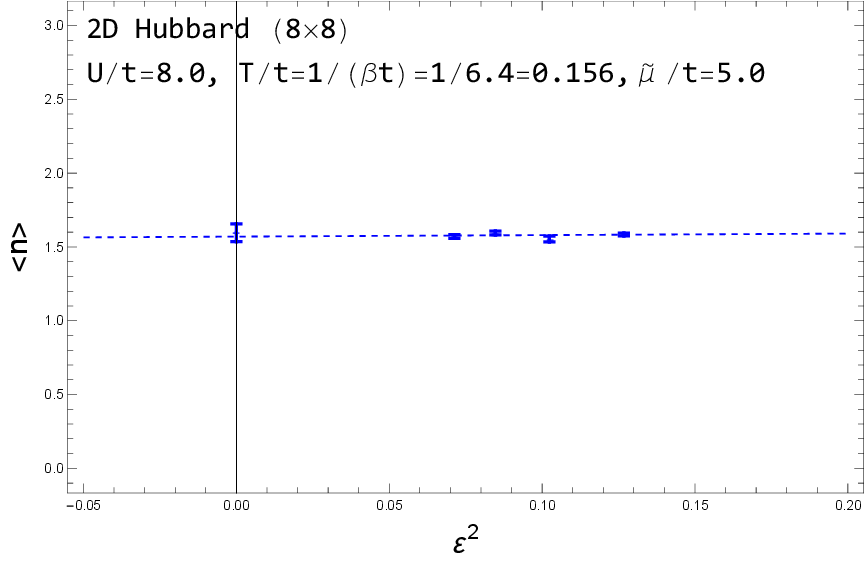}
  \hspace{1mm}
  \includegraphics[width=52mm]{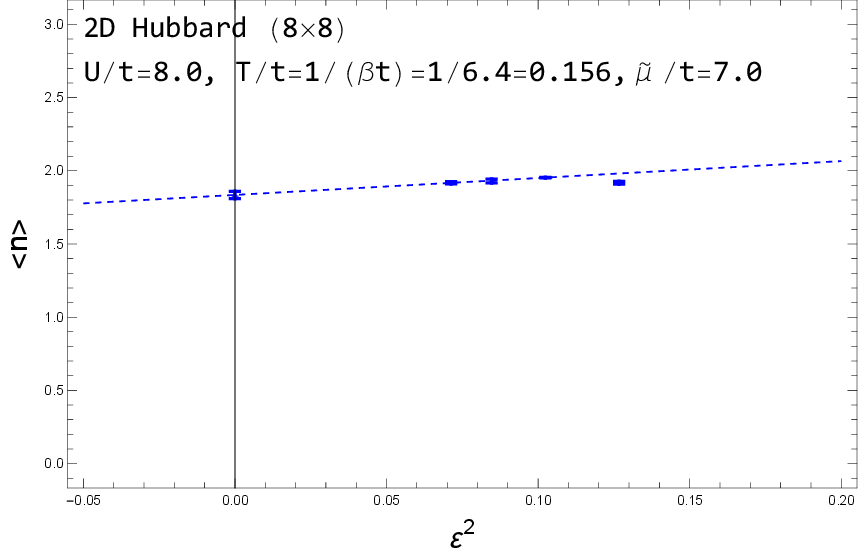}
  \vspace{3ex}
  \\
  \includegraphics[width=52mm]{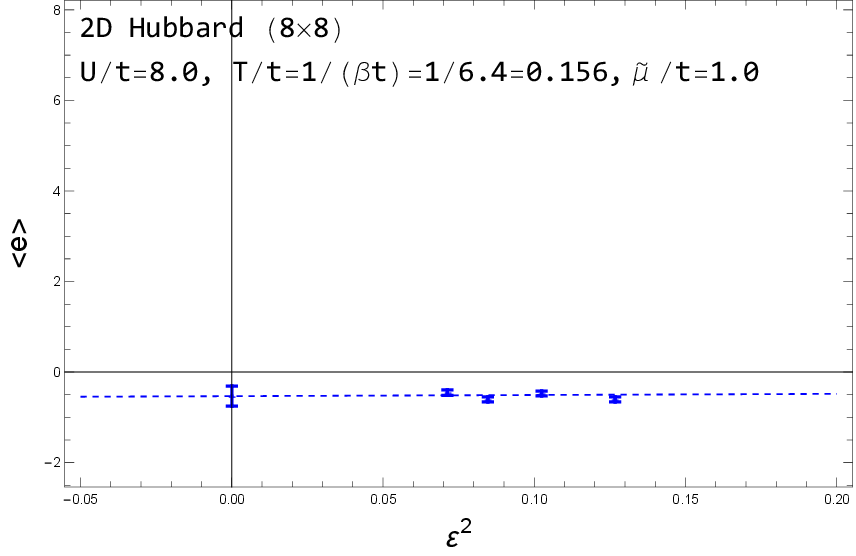}
  \hspace{1mm}
  \includegraphics[width=52mm]{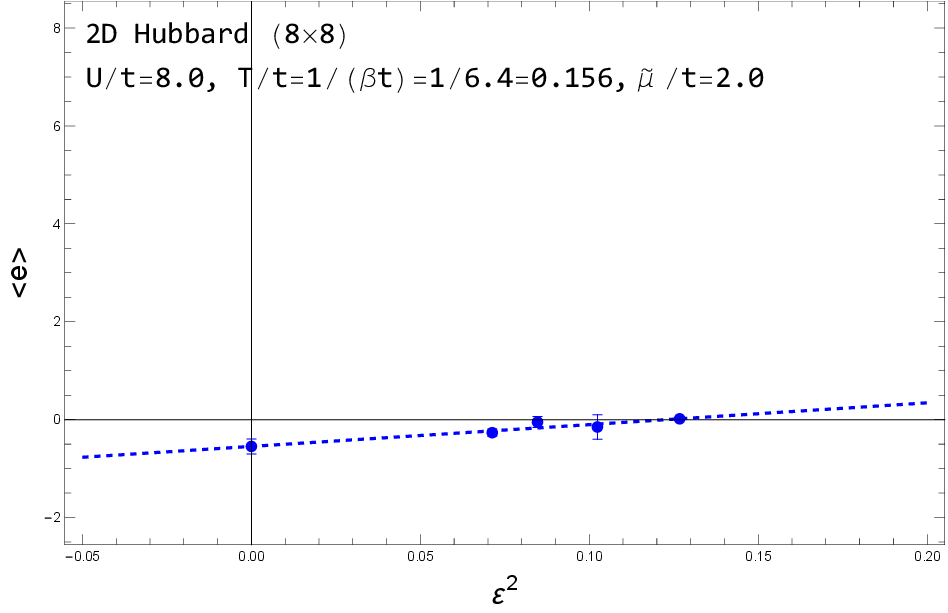}
  \hspace{1mm}
  \includegraphics[width=52mm]{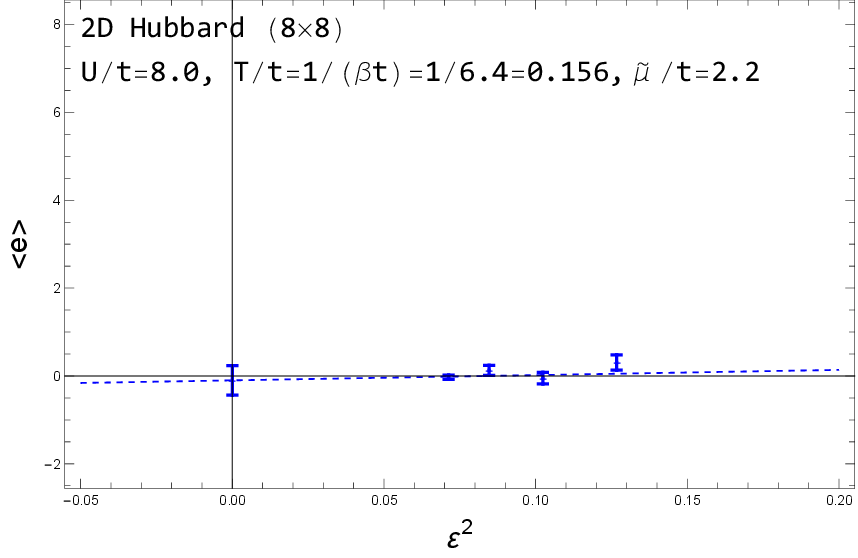}
  \vspace{3ex}
  \\
  \includegraphics[width=52mm]{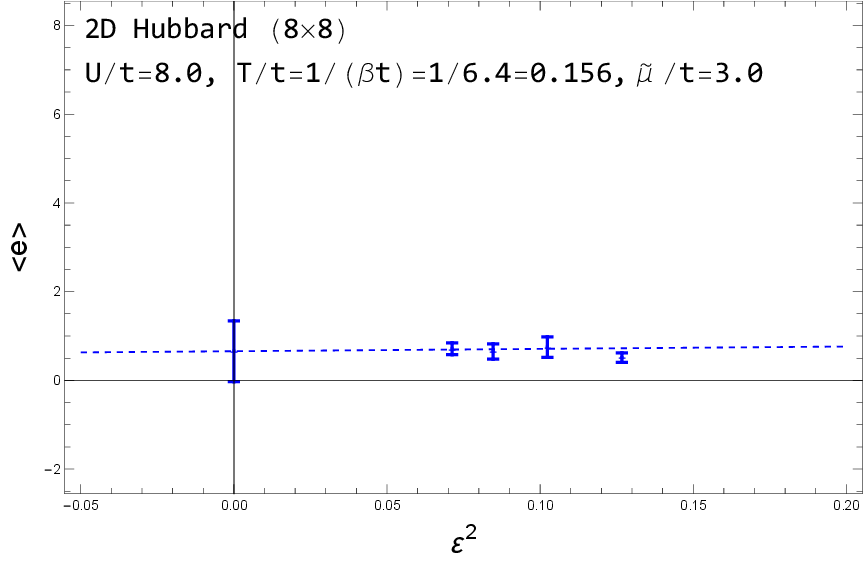}
  \hspace{1mm}
  \includegraphics[width=52mm]{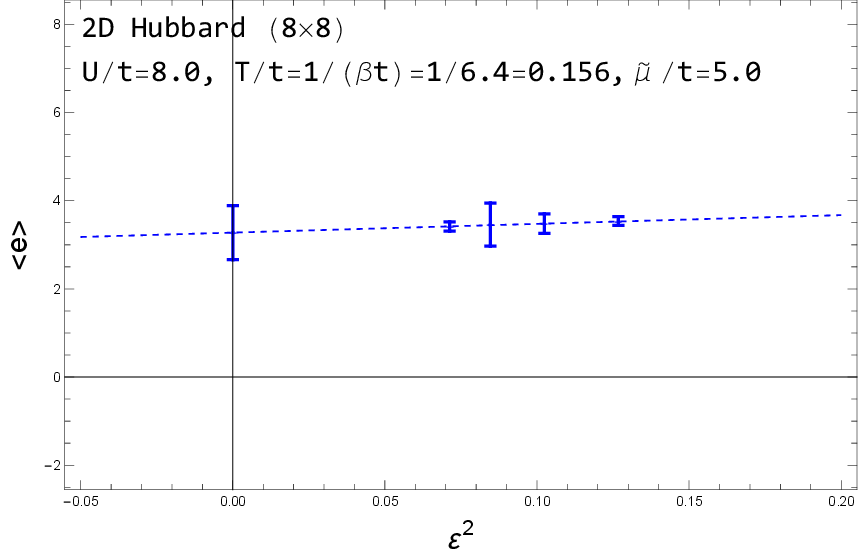}
  \hspace{1mm}
  \includegraphics[width=52mm]{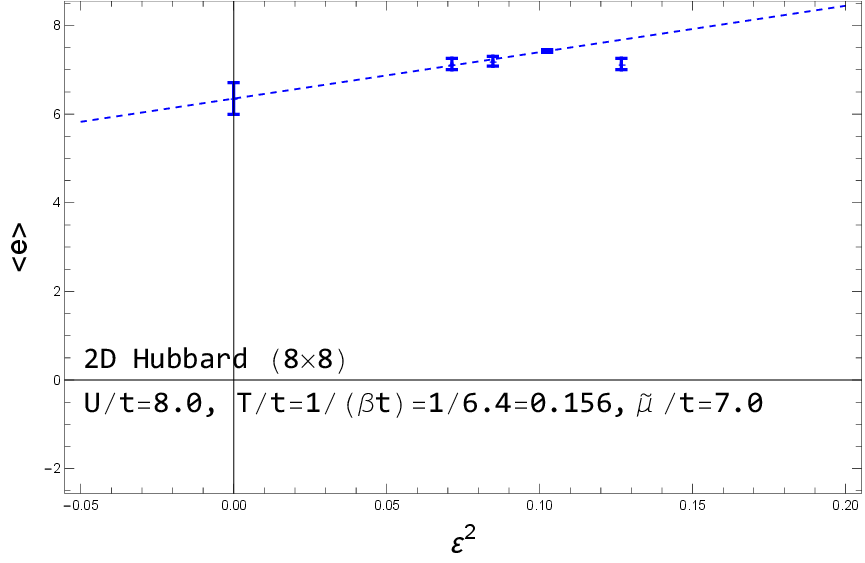}
  \caption{%
    Finite-$\epsilon$ effects 
    in the number density $\vev{n}$ (top 6 panels) 
    and the energy density $\vev{e}$ (bottom 6 panels) 
    obtained using the combined WV-HMC algorithm 
    for $\tilde\mu = 1.0,\,2.0,\,2.2,\,3.0,\,5.0,\,7.0$.
  }
\label{fig:observables_8x8_eps_scaling}
\end{figure}

\begin{figure}[ht]
  \centering
  \includegraphics[width=140mm]
    {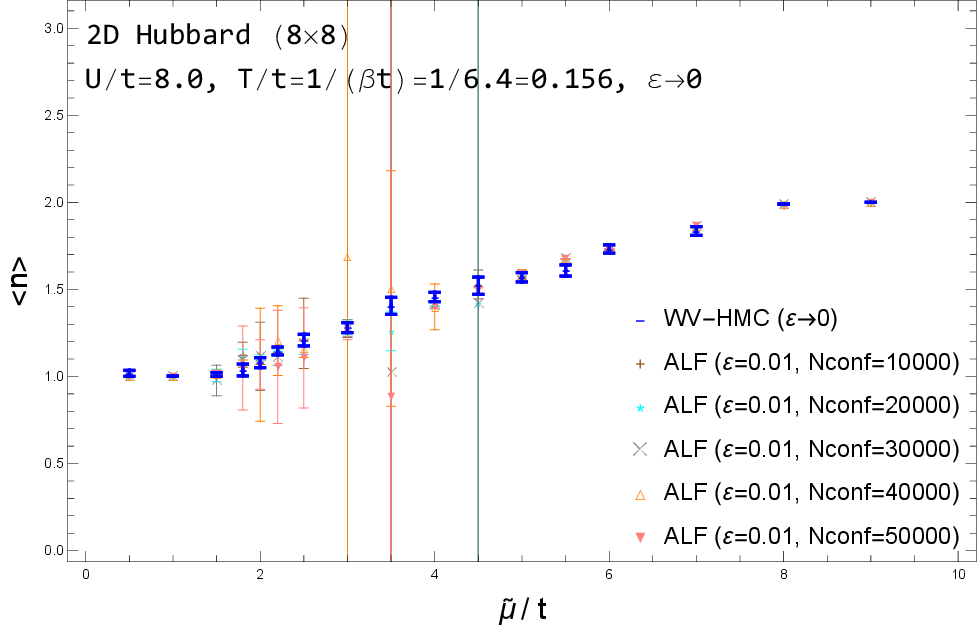}
  \vspace{3ex}\\
  \includegraphics[width=140mm]
    {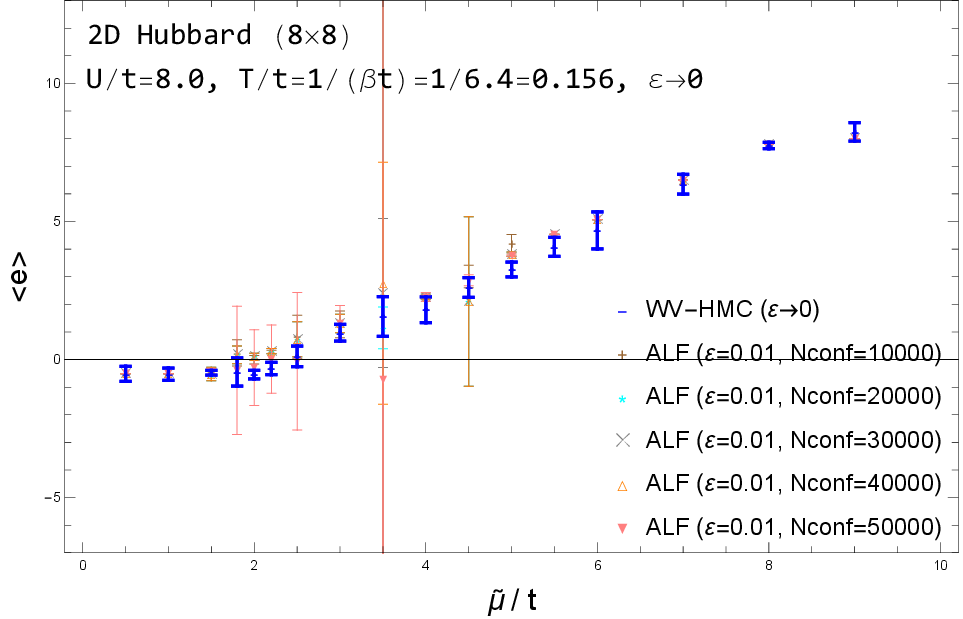}
  \caption{%
    Number density $\vev{n}$ and energy density $\vev{e}$ 
    in the continuum limit $\epsilon \to 0$ 
    obtained using the combined WV-HMC algorithm, 
    compared with results from ALF.
  }
\label{fig:observables_8x8_wv_vs_alf}
\end{figure}

Figure~\ref{fig:observables_8x8_eps_scaling} shows that 
the finite-$\epsilon$ effects in the observables scale as $O(\epsilon^2)$, 
in agreement with the expectation from Eqs.~\eqref{scaling_n} and \eqref{scaling_e}. 
To obtain the limit $\epsilon \to 0$, 
we perform a $\chi^2$ fit 
using a linear ansatz in $\epsilon^2$ 
for the results at $\epsilon = 0.27,\,0.29,\,0.32$ 
(as well as the result at $\epsilon = 0.36$ 
when it is consistent with the linear trend).%
\footnote{ 
  The sample sizes at $\epsilon=0.27, 0.29, 0.32, 0.36$
  are 12--22, 20--26, 30--42, 20--32 (varying depending on $\tilde \mu$), 
  respectively. 
  The necessary bin sizes at the blocked jackknife method are around three.
} 
The obtained results are shown in Fig.~\ref{fig:observables_8x8_wv_vs_alf}, 
where those from ALF at $\epsilon = 0.01$ are also displayed for comparison.%
\footnote{ 
  The ALF results can be considered to have reached the continuum limit 
  because they no longer vary for $\epsilon < 0.01$.
} 
We find that WV-HMC yields observables with controlled statistical errors 
even in the continuum limit.%
\footnote{ 
  In Ref.~\cite{Fukuma:2025uzg}, 
  we compared WV-HMC results at $\epsilon = 0.32$ 
  with ALF results at $\epsilon = 0.01$, 
  and found significant discrepancies 
  in a parameter region where the sign problem is not serious 
  (and thus ALF results are reliable). 
  The continuum limit of WV-HMC now yields results consistent with ALF, 
  indicating that these discrepancies were due to finite-$\epsilon$ effects.
} 
%

\section{Conclusions and outlook}
\label{sec:conclusions}

In this paper, 
we considered cases 
where the maximum flow time in WV-HMC can be set to a small value, 
as realized in the doped Hubbard model 
by using a redundant, nonphysical parameter $\alpha$ 
introduced in Ref.~\cite{Beyl:2017kwp}. 
In these cases, however, 
the worldvolume becomes a thin layer, 
and ergodicity issues may arise 
due to the inefficient exploration of this space. 
To address these issues, 
we proposed embedding GT-HMC into the WV-HMC framework 
and provided a mathematical proof of this embedding. 
The key idea behind this construction is 
that the tangent bundle $T\Sigma_t$ used in GT-HMC 
can be symplectically embedded 
into the tangent bundle $T\calR$ used in WV-HMC. 

We applied the combined algorithm to the two-dimensional doped Hubbard model 
on an $8 \times 8$ lattice at $T/t = 1/6.4 \simeq 0.156$ and $U/t = 8.0$. 
We demonstrated that the pure and combined algorithms 
yield consistent results within statistical errors 
at Trotter step $\epsilon = 0.32$. 
Furthermore, 
taking advantage of the ability of the combined algorithm 
to simulate larger spacetime lattices, 
we took the zero Trotter step limit ($\epsilon \to 0$) 
using the results obtained at $\epsilon = 0.27,\,0.29,\,0.32,\,0.36$ 
(corresponding to temporal sizes $N_t = 24,\,22,\,20,\,18$). 
Even though the sample sizes are still small, 
we confirmed that the continuum limits are obtained 
for the entire range of the chemical potential 
with controlled statistical errors.

Although it is straightforward to keep increasing the statistics 
or go to larger spacetime lattices, 
the current algorithm based on direct solvers 
for the inversion of the fermion matrices $D_{a/b}$, 
which scales as $O(N^3)$, is computationally expensive. 
There is indeed another approach, 
which uses pseudofermions and iterative solvers 
and is anticipated to reduce the computational cost to $O(N^2)$, 
although it requires careful tuning of simulation parameters 
\cite{Fukuma:2025esu}. 
The embedding prescription presented in this work 
is expected to remain applicable in that setting as well. 
Research in this direction is currently in progress 
and will be reported elsewhere.

\section*{Acknowledgments}

The authors thank 
Sinya Aoki, Fakher F.~Assaad, Masatoshi Imada, 
Ken-Ichi Ishikawa, Issaku Kanamori, Yoshio Kikukawa, 
Nobuyuki Matsumoto, Yusuke Nomura, Maksim Ulybyshev, 
Youhei Yamaji, and Shiwei Zhang 
for valuable discussions. 
M.F.\ acknowledges that 
parts of the basic lattice-field infrastructure of the code 
were developed using the FermiQCD/MDP library \cite{DiPierro:2001yu}. 
The WV-HMC algorithmic components, 
together with the configuration generation, 
random-number generation, 
fermion-matrix construction and solvers, 
and routines for observable measurements, 
were implemented independently by M.F.
This work was partially supported by JSPS KAKENHI
(Grant Numbers JP20H01900, JP21K03553, JP23H00112, JP23H04506, JP24K07052, JP25H01533);
by MEXT as 
``Program for Promoting Researches on the Supercomputer Fugaku'' 
(Simulation for basic science: approaching the new quantum era,
JPMXP1020230411);
and by SPIRIT2 2025 of Kyoto University.




\baselineskip=0.9\normalbaselineskip




\begin{thebibliography}{99}
  
  
\bibitem{Witten:2010cx}
E.~Witten,
``Analytic continuation of Chern-Simons theory,''
AMS/IP Stud.\ Adv.\ Math.\ \textbf{50}, 347-446 (2011)
[arXiv:1001.2933 [hep-th]].
  
\bibitem{Cristoforetti:2012su} 
M.~Cristoforetti, F.~Di Renzo and L.~Scorzato,
``New approach to the sign problem in quantum field theories: 
High density QCD on a Lefschetz thimble,''
Phys.\ Rev.\ D {\bf 86}, 074506 (2012)
[arXiv:1205.3996 [hep-lat]].
  
\bibitem{Cristoforetti:2013wha} 
M.~Cristoforetti, F.~Di Renzo, A.~Mukherjee and L.~Scorzato,
``Monte Carlo simulations on the Lefschetz thimble: Taming the sign problem,''
Phys.\ Rev.\ D {\bf 88}, no.\ 5, 051501(R) (2013)
[arXiv:1303.7204 [hep-lat]].
  
\bibitem{Fujii:2013sra} 
H.~Fujii, D.~Honda, M.~Kato, Y.~Kikukawa, S.~Komatsu and T.~Sano,
``Hybrid Monte Carlo on Lefschetz thimbles - A study of the residual sign problem,''
JHEP {\bf 1310}, 147 (2013)
[arXiv:1309.4371 [hep-lat]].
  
\bibitem{Fujii:2015bua}
H.~Fujii, S.~Kamata and Y.~Kikukawa,
``Lefschetz thimble structure 
in one-dimensional lattice Thirring model at finite density,''
JHEP \textbf{11}, 078 (2015)
[erratum: JHEP \textbf{02}, 036 (2016)]
[arXiv:1509.08176 [hep-lat]].
  
\bibitem{Fujii:2015vha}
H.~Fujii, S.~Kamata and Y.~Kikukawa,
``Monte Carlo study of Lefschetz thimble structure 
in one-dimensional Thirring model at finite density,''
JHEP \textbf{12}, 125 (2015)
[erratum: JHEP \textbf{09}, 172 (2016)]
[arXiv:1509.09141 [hep-lat]].
  
\bibitem{Alexandru:2015xva} 
A.~Alexandru, G.~Ba\c sar and P.~Bedaque,
``Monte Carlo algorithm for simulating fermions on Lefschetz thimbles,''
Phys.\ Rev.\ D {\bf 93}, no.\ 1, 014504 (2016)
[arXiv:1510.03258 [hep-lat]].
  
\bibitem{Alexandru:2015sua} 
A.~Alexandru, G.~Ba\c sar, P.~F.~Bedaque, G.~W.~Ridgway and N.~C.~Warrington,
``Sign problem and Monte Carlo calculations beyond Lefschetz thimbles,''
JHEP {\bf 1605}, 053 (2016)
[arXiv:1512.08764 [hep-lat]].
  
\bibitem{Alexandru:2017lqr}
A.~Alexandru, G.~Basar, P.~F.~Bedaque and G.~W.~Ridgway,
``Schwinger-Keldysh formalism on the lattice: A faster algorithm and its application to field theory,''
Phys. Rev. D \textbf{95}, no.11, 114501 (2017)
[arXiv:1704.06404 [hep-lat]].

\bibitem{Fukuma:2017fjq} 
M.~Fukuma and N.~Umeda,
``Parallel tempering algorithm for integration over Lefschetz thimbles,''
PTEP {\bf 2017}, no.\ 7, 073B01 (2017)
[arXiv:1703.00861 [hep-lat]].
  
\bibitem{Alexandru:2017oyw} 
A.~Alexandru, G.~Ba\c sar, P.~F.~Bedaque and N.~C.~Warrington,
``Tempered transitions between thimbles,''
Phys.\ Rev.\ D {\bf 96}, no.\ 3, 034513 (2017)
[arXiv:1703.02414 [hep-lat]].
  
\bibitem{Fukuma:2019wbv}
M.~Fukuma, N.~Matsumoto and N.~Umeda,
``Applying the tempered Lefschetz thimble method 
to the Hubbard model away from half-filling,''
Phys. Rev. D \textbf{100}, no.11, 114510 (2019)
[arXiv:1906.04243 [cond-mat.str-el]].
  
\bibitem{Fukuma:2020fez}
M.~Fukuma and N.~Matsumoto,
``Worldvolume approach to the tempered Lefschetz thimble method,''
PTEP \textbf{2021}, no.2, 023B08 (2021)
[arXiv:2012.08468 [hep-lat]].
  
\bibitem{Fukuma:2021aoo}
M.~Fukuma, N.~Matsumoto and Y.~Namekawa,
``Statistical analysis method for the worldvolume hybrid Monte Carlo algorithm,''
PTEP \textbf{2021}, no.12, 123B02 (2021)
[arXiv:2107.06858 [hep-lat]].
  
\bibitem{Fukuma:2023eru}
M.~Fukuma,
``Simplified Algorithm for the Worldvolume HMC and the Generalized Thimble HMC,''
PTEP \textbf{2024}, no.5, 053B02 (2024)
[arXiv:2311.10663 [hep-lat]].
  
\bibitem{Fukuma:2025gya}
M.~Fukuma,
``Worldvolume Hybrid Monte Carlo algorithm for group manifolds,''
[arXiv:2506.12002 [hep-lat]].
  
\bibitem{Fukuma:2025uzg}
M.~Fukuma and Y.~Namekawa,
``Applying the Worldvolume Hybrid Monte Carlo method to the Hubbard model away from half filling,''
[arXiv:2507.23748 [cond-mat.str-el]].

\bibitem{Namekawa:2024ert}
M.~Fukuma and Y.~Namekawa,
``Applying the Worldvolume Hybrid Monte Carlo method 
to the finite-density complex $\phi^4$ model and the Hubbard model,''
PoS \textbf{LATTICE2023}, 178 (2024)
  
\bibitem{Fukuma:2025esu}
M.~Fukuma and Y.~Namekawa,
``Applying the Worldvolume Hybrid Monte Carlo method 
to the two-dimensional Hubbard model,''
PoS \textbf{LATTICE2024}, 053 (2025)
  
\bibitem{Andersen:1983}
H.~C.~Andersen,
``RATTLE: A ``velocity'' version of the SHAKE algorithm 
for molecular dynamics calculations,''
J.\ Comput.\ Phys.\ {\bf 52}, 24 (1983).

\bibitem{Leimkuhler:1994} 
B.~J.~Leimkuhler and R.~D.~Skeel, 
``Symplectic numerical integrators in constrained Hamiltonian systems,''
J.\ Comput.\ Phys.\ {\bf 112}, 117 (1994).

\bibitem{Beyl:2017kwp}
S.~Beyl, F.~Goth and F.~F.~Assaad,
``Revisiting the Hybrid Quantum Monte Carlo Method 
for Hubbard and Electron-Phonon Models,''
Phys. Rev. B \textbf{97}, no.8, 085144 (2018)
[arXiv:1708.03661 [cond-mat.str-el]].
  
\bibitem{Mukherjee:2014hsa}
A.~Mukherjee and M.~Cristoforetti,
``Lefschetz thimble Monte Carlo for many-body theories: A Hubbard model study,''
Phys. Rev. B \textbf{90}, no.3, 035134 (2014)
[arXiv:1403.5680 [cond-mat.str-el]].
  
\bibitem{Ulybyshev:2017hbs}
M.~V.~Ulybyshev and S.~N.~Valgushev,
``Path integral representation for the Hubbard model 
with reduced number of Lefschetz thimbles,''
[arXiv:1712.02188 [cond-mat.str-el]].
  
\bibitem{Ulybyshev:2019hfm}
M.~Ulybyshev, C.~Winterowd and S.~Zafeiropoulos,
``Taming the sign problem of the finite density Hubbard model 
via Lefschetz thimbles,''
[arXiv:1906.02726 [cond-mat.str-el]].
  
\bibitem{Ulybyshev:2019fte}
M.~Ulybyshev, C.~Winterowd and S.~Zafeiropoulos,
``Lefschetz thimbles decomposition for the Hubbard model on the hexagonal lattice,''
Phys. Rev. D \textbf{101}, no.1, 014508 (2020)
[arXiv:1906.07678 [cond-mat.str-el]].
  
\bibitem{Ulybyshev:2022kxq}
M.~Ulybyshev, C.~Winterowd, F.~Assaad and S.~Zafeiropoulos,
``Instanton gas approach to the Hubbard model,''
Phys. Rev. B \textbf{107}, no.4, 045143 (2023)
[arXiv:2207.06297 [cond-mat.str-el]].
  
\bibitem{Ulybyshev:2024kdr}
M.~Ulybyshev and F.~F.~Assaad,
``Beyond the instanton gas approach: dominant thimbles approximation 
for the Hubbard model,''
[arXiv:2407.09452 [cond-mat.str-el]].
  
\bibitem{Alexandru:2019} 
A.~Alexandru, 
``Improved algorithms for generalized thimble method,''
talk at the 37th international conference on lattice field theory, Wuhan, 2019. 
  
\bibitem{Fukuma:2019uot}
M.~Fukuma, N.~Matsumoto and N.~Umeda,
``Implementation of the HMC algorithm on the tempered Lefschetz thimble method,''
[arXiv:1912.13303 [hep-lat]].
  
\bibitem{Bercx:2017pit}
M.~Bercx, F.~Goth, J.~S.~Hofmann and F.~F.~Assaad,
``The ALF (Algorithms for Lattice Fermions) project release 1.0. 
Documentation for the auxiliary field quantum Monte Carlo code,''
SciPost Phys. \textbf{3}, no.2, 013 (2017)
[arXiv:1704.00131 [cond-mat.str-el]].
  
\bibitem{ALF:2020tyi}
F.~F.~Assaad \textit{et al.} [ALF],
``The ALF (Algorithms for Lattice Fermions) project release 2.4. 
Documentation for the auxiliary-field quantum Monte Carlo code,''
SciPost Phys. Codeb. \textbf{2022}, 1 (2022)
[arXiv:2012.11914 [cond-mat.str-el]].
   
\bibitem{DiPierro:2001yu}
 M.~Di Pierro,
``FermiQCD: A Tool Kit for Parallel Lattice QCD Applications,''
 Nucl. Phys. B Proc. Suppl. \textbf{106}, 1034-1036 (2002)
 [arXiv:hep-lat/0110116 [hep-lat]].

\end{thebibliography}
\end{document}